\documentclass[twocolumn,twocolappendix]{aastex631}
\usepackage{amsmath,mathtools,mathrsfs}
\usepackage{graphicx}
\usepackage{dcolumn}
\usepackage{bm}
\usepackage{accents}

\usepackage{ulem}
\usepackage{xcolor}
\usepackage{amsmath}

\usepackage{booktabs}
\usepackage{multirow}
\usepackage{enumitem} 

\DeclareFontFamily{OT1}{pzc}{}
\DeclareFontShape{OT1}{pzc}{m}{it}{<-> s * [1.10] pzcmi7t}{}
\DeclareMathAlphabet{\mathpzc}{OT1}{pzc}{m}{it}

\defcitealias{Ozel2021}{Paper~I}

\usepackage{graphicx}
\usepackage[percent]{overpic} 
\usepackage{dcolumn}
\usepackage{bm}
\usepackage{accents}

\graphicspath{{Figures/}}

\usepackage{ulem}
\usepackage{xcolor}
\usepackage{xspace}
\usepackage{amsmath}

\newcommand{\sub}[1]{\rm \scriptscriptstyle {#1}} 
\newcommand{\nicevec}[1]{{\hspace*{-2pt}\vec{\kern2pt {#1}}}} 
\newcommand{\rg}[0]{{\rm r}_{\sub g}} 

\shorttitle{Black hole images as tests of GR: effects of spacetime geometry}
\shortauthors{Younsi, Psaltis \& \"Ozel}

\begin{document}

\title{Black Hole Images as Tests of General Relativity: Effects of Spacetime Geometry}


\author[0000-0001-9283-1191]{Ziri Younsi}
\altaffiliation{UKRI Stephen Hawking Fellow}
\affiliation{Mullard Space Science Laboratory, University College London, Holmbury St.~Mary, Dorking, Surrey, RH5 6NT, UK}

\author[0000-0003-1035-3240]{Dimitrios Psaltis}
\affiliation{Department of Astronomy and Steward Observatory, University of Arizona, 933 N.~Cherry Ave., Tucson, AZ 85721, USA}

\author[0000-0003-4413-1523]{Feryal \"Ozel}
\affiliation{Department of Astronomy and Steward Observatory, University of Arizona, 933 N.~Cherry Ave., Tucson, AZ 85721, USA}

\begin{abstract}
The images of supermassive black holes surrounded by optically-thin, radiatively-inefficient accretion flows, like those observed with the Event Horizon Telescope, are characterized by a bright ring of emission surrounding the black-hole shadow.
In the Kerr spacetime this bright ring, when narrow, closely traces the boundary of the shadow and can, with appropriate calibration, serve as its proxy.
The present paper expands the validity of this statement by considering two particular spacetime geometries: a solution to the field equations of a modified gravity theory and another that parametrically deviates from Kerr but recovers the Kerr spacetime when its deviation parameters vanish.
A covariant, axisymmetric analytic model of the accretion flow based on conservation laws and spanning a broad range of plasma conditions is utilized to calculate synthetic non-Kerr black-hole images, which are then analysed and characterized. We find that in all spacetimes: {\it (i)} it is the gravitationally-lensed unstable photon orbit that plays the critical role in establishing the diameter of the rings observed in black-hole images, not the event horizon or the innermost stable circular orbit, {\it (ii)} bright rings in these images scale in size with, and encompass, the boundaries of the black-hole shadows, even when deviating significantly from Kerr, and {\it (iii)} uncertainties in the physical properties of the accreting plasma introduce subdominant corrections to the relation between the diameter of the image and the diameter of the black-hole shadow.
These results provide important new theoretical justification for using black-hole images to probe and test the spacetimes of supermassive black holes.
\end{abstract}


\section{Introduction}
The horizon-scale images obtained with the Event Horizon Telescope (EHT) of the black hole in the center of M87 have opened a new avenue for probing spacetimes of black holes and testing the theory of General Relativity in the strong-field regime~\citep{PaperI,EHTC2019_PaperVI,Psaltis2020,Kocherlakota2021}.
The images are characterized by a deep central brightness depression, which has been identified with the black-hole shadow, surrounded by a bright ring of emission produced by radiation emerging from the accreting plasma.
This radiation, which is produced in the vicinity of the event horizon, is subject to strong gravitational lensing as it propagates through the black-hole spacetime, i.e., influenced by the geometrical structure of the background gravitational field of the black hole. Using black-hole images to infer spacetime properties requires establishing a connection between this bright ring-like structure and the various characteristic properties of the spacetime.

In any black-hole spacetime, there are well defined characteristic radii, such as the radius of the event horizon, the radii of spherical photon orbits, and the radius of the innermost stable circular orbit, which is straightforward to calculate mathematically\footnote{Not all spacetimes possess all of these characteristic radii, even within General Relativity. For example, naked singularities do not have horizons and a number of them do not have unstable circular orbits either (see, e.g.,~\citealt{Gair2008}).}~\citep{Bardeen1972}.
In recent years, characteristic radii have been calculated for a large number of non-Kerr spacetimes that are solutions to various modifications of General Relativity~(see, e.g., \citealt{Bambi2010,Amarilla2010,Amarilla2012,Amarilla2013,Abdu2013,Ayzenberg2014,Sakai2014,Tsukamoto2014,Cunha2015,Cunha2017a,Moffat2015,Chiba2017,Ghasemi2020,Kumar2020b,Tsupko2020,Xavier2020,Fathi2021,Li2021}).
All of these radii determine the trajectories of the plasma fluid elements (particles) and those of the photons (radiation) emitted by the plasma and, in principle, can affect the resulting images.
However, when using black-hole images to carry out precise tests of gravity, several additional questions arise.
Which aspects of the spacetime play the most critical role in image formation?
Is there a predictable relation between the bright emission ring that is observable and the characteristic radii of the spacetime?
And finally, to what extent do plasma processes complicate this aforementioned relation?

The image of a Kerr black hole that is surrounded by an optically thin plasma is characterized by an abrupt drop in brightness which marks the boundary of the so-called black-hole shadow~\citep{Bardeen1973}.
This boundary occurs at the gravitationally-lensed image of the photon orbits near the horizon.
Its size and shape depend very weakly on the spin of the black hole and the inclination of the observer because of a fortuitous near cancellation of the effects of frame dragging and of the spacetime quadrupole~\citep{Johannsen2010b}.
The image brightness at this boundary is formally infinite, forming a very narrow but bright photon ring.
However, the presence of an extended plasma distribution around the black hole generates a broader ring of emission on the images, which is what can be resolved with the EHT~\citep{PaperV}.

Since the advent of the first EHT image of M87*, and most recently of Sagittarius A* \citep{EHTPaperI_2022}, several authors have attempted to address different aspects of the above questions (e.g., \citealt[]{Volkel2021,Glampedakis2021,Lara2021,Gralla2021,Kocherlakota2022}; see \citealt[hereafter Paper~I,]{Ozel2021} for a detailed discussion).
However, the choices of plasma effects employed in these studies have relied either on simplified models of the radiating plasma that utilize radial profiles for the key thermodynamic parameters, or on even simpler toy models. More importantly, several of these studies introduced artificial cut-offs in the plasma emissivity profiles close to the black-hole horizon at some ad hoc radius such as that of the innermost stable circular orbit. As discussed in Paper I, such constructions do not satisfy fundamental conservation laws.

A covariant model of the accreting plasma is, therefore, essential for constructing reliable synthetic images of non-Kerr black holes and for maintaining the generality and validity of the results. In essence, one must explore both the black hole spacetime geometry and the surrounding plasma properties when seeking to establish whether there exists a quantifiable connection between persistent image features and key properties of the underlying spacetime geometry.

In Paper~I we employed covariant analytic models to show that the bright ring in black hole images is coupled to and encompasses the photon ring in the Kerr spacetime for any accretion flow that satisfies conservation laws and basic thermodynamic principles.
In this paper, we extend the investigation to non-Kerr spacetimes, addressing the questions posed above.
In order to explore different properties of non-Kerr spacetimes and ensure the generality of our results, we utilize two classes of non-Kerr spacetimes: the first represents known solutions to modified gravity theories; the second represents spacetimes designed to be parametrically different from the Kerr metric without the requirement that they are solutions to any particular field equations.

For the first class, we employ the time-independent, axisymmetric Einstein-Maxwell-Dilaton-Axion metric, which is a known solution to the field equations that arise from the 4D compactification and low-energy truncation of heterotic string theories~\citep{Garcia1995}.
In addition to the metric, the field equations involve an electromagnetic field, a dilaton scalar field, and an axion field.
Taking different limits of this general metric reproduces other known solutions, such as the Sen metric~\citep{Sen1992}.
The characteristic radii for variants of this spacetime and the boundaries of black-hole shadows have been explored previously and found to depend on the parameters that control the couplings of the additional fields~\citep{Wei2013,Younsi2016}.

For the second class, we employ the Johannsen-Psaltis metric, which is a time-independent, axisymmetric metric that has been parametrically modified away from Kerr~\citep{Johannsen2011,Johannsen2013b}.
The modification has been imposed in such a way that the metric remains pathology-free outside the horizon, while admitting a Carter-like integral of motion for a broad range of the deviation parameters (see, e.g., \citealt{Johannsen2013}).
The characteristic radii and shadow boundaries for this metric have also been explored previously~\citep{Johannsen2013c,Medeiros2020} and, in fact, used to place constraints on modifications of the Kerr metric based on EHT images~\citep{Psaltis2020}.

In order to go beyond the simple mathematical descriptions of shadow boundaries explored in earlier papers, we utilize these two spacetime metrics to calculate images and brightness profiles using both a spherically symmetric simple emissivity profile as well as the full accretion plasma model we developed in \citetalias{Ozel2021}.
This model is a covariant, semi-analytic solution to the set of basic conservation laws which govern the dynamics and thermodynamics of the gas accreting onto the black hole.
It incorporates parameters that can be modified in order to capture, e.g., different heating rates or efficiencies of angular momentum transport in the flow. 
The model has been calibrated against time-dependent general-relativistic magneto-hydrodynamics (GRMHD) simulations but allows for a much broader exploration of physical conditions in the accretion flow than such simulations currently permit.

We use these models to investigate whether the relationship between the basic features of optically thin black-hole images like those observed with the EHT, and spacetime characteristics, remain qualitatively unchanged even in non-Kerr spacetimes.
This paper is organised as follows.
In Section~\ref{sec:metrics}, we detail the different spacetime geometries used in this study and provide an overview of the importance of charcateristic radii and the procedure for their evaluation.
In Section~\ref{sec:plasma}, we present a simple toy plasma model with free-falling plasma velocities and subsequently summarize the full covariant plasma model and generalized plasma velocity profile employed throughout the bulk of calculations in this study.

We present in Sec.~\ref{sec:results} the results of varying spacetime geometry on black hole images, first for the toy model and subsequently for the full covariant plasma model.
Following this, we present the results of an exploration of over $10^{5}$ non-Kerr images, probing a broad range of both spacetime parameters and plasma properties, finding that, as in the Kerr metric, the significant majority of non-Kerr model images possess bright rings which are slightly larger than the black hole shadow.
We also present the results from a study of black holes with extreme deviation parameters, yielding shadows between $2.5$ and $1000$ times larger than is possible in the Kerr spacetime, finding the above conclusions remain unchanged.
Finally, in Section~\ref{Sec:Discussion_and_Conclusions}, we present the conclusions and discussion.
Several appendices provide further detail on the calculation procedure for characteristic radii and comparison with previous known results (Appendix \ref{Appendix_Char_Radii}), non-Kerr plasma free-fall velocities not previously published in the literature (Appendix \ref{Appendix:free_fall}), example images excluded from the analysis (Appendix \ref{Sec:Appendix_Library}), and a cross-validation of non-Kerr covariant radiation transport codes (Appendix \ref{Sec:GRRT_Code_Comparison}).       
\section{Spacetime Geometries}
\label{sec:metrics}
In this section, we provide the details of the three spacetime metrics we employ in this study.
Hereafter we adopt the $[-,+,+,+]$ signature convention.
When specifying tensors, Greek indices (e.g., $\mu,\,\nu$) span $\left(0,1,2,3\right)$ and Latin indices (e.g., $i,\,j$) span $\left(1,2,3\right)$. In this study $\left(0,1,2,3\right)$ correspond, respectively, to the coordinates $\left(t,r,\theta,\phi\right)$.
\newline
\subsection{The Kerr metric}
We begin with the Kerr metric \citep{Kerr1963}, which is considered the most astrophysically relevant black hole solution and describes the exterior of a static and axisymmetric black hole in General Relativity (GR).
It also serves as the reference solution, to which we compare our non-GR black hole solutions.
In Boyer-Lindquist (oblate spheroidal) coordinates \citep{Boyer1967}, the Kerr metric line element is given by:
\begin{equation}
\begin{aligned}
{\rm d}s^{2} = &-\left(1 - \frac{2\,\rg \, r}{\Sigma}\right) \, c^{2}{\rm d}t^{2} - \frac{4a\,\rg \, r\sin^2\theta}{\Sigma} \, c\,{\rm d}t \, {\rm d}\phi \\
&+ \frac{\Sigma}{\Delta} \, {\rm d}r^{2} + \Sigma \, {\rm d}\theta^{2} + \frac{\mathcal{A} \sin^{2} \theta}{\Sigma} \, {\rm d}\phi^{2} \,,
\end{aligned}
\end{equation}
where
\begin{subequations}
\begin{eqnarray}
\Sigma &:=& r^{2} + a^{2} \cos^{2} \theta \,, \\
\Delta &:=& r^{2} - 2\,\rg \, r + a^{2} \,, \\
\mathcal{A} &:=& \left(r^{2} + a^{2}\right)^{2} - a^{2}\Delta\sin^{2}\theta \,.
\end{eqnarray}
\end{subequations}
Herein, $\rg\equiv GMc^{-2}$ denotes the gravitational radius of the black hole, where $M$ is its mass, and $G$ and $c$ denote Newton's gravitational constant and the speed of light, respectively.
Furthermore, $a \equiv J/(c M)$ denotes the black hole's {\it dimensional} spin parameter (units of length) and $J$ denotes its total angular momentum.
The {\it dimensionless} spin parameter may then be defined as $a_{\sub *} \equiv cJ/(GM^{2}) \equiv a/{\rm r}_{\sub g}$.
The numerical calculations in this study adopt the geometrical unit convention (wherein $G=c=1$) and also let $M=1$, which is equivalent to normalizing all length scales to units of $\rg$.
\subsection{The Einstein-Maxwell-Dilaton-Axion metric}
The second black hole solution we consider is the Einstein-Maxwell-Dilaton-Axion (EMDA) metric.
It is chosen to serve as a particular, demonstrative non-GR black hole solution, with specific metric parameters corresponding to physical field couplings.
Following \cite{Garcia1995}, the EMDA line element for a static, axisymmetric black hole may be written as:
\begin{equation}
\begin{aligned}
{\rm d}s^{2} = &-\left( \frac{\widehat{\Delta} - a^{2}\sin^{2}\theta}{\widehat{\Sigma}} \right)c^{2} {\rm d}t^{2} \\
&-\frac{2a \left(\delta - \widehat{\Delta} W\right)\sin^{2}\theta}{\widehat{\Sigma}}\, c\,{\rm d}t \, {\rm d}\phi + \frac{\widehat{\Sigma}}{\widehat{\Delta}} \, {\rm d}r^{2} \\
&+ \widehat{\Sigma} \, {\rm d}\theta^{2} + \frac{\widehat{\mathcal{A}}\sin^{2}\theta}{\widehat{\Sigma}} \, {\rm d}\phi^{2} \,,
\label{EMDA_new}
\end{aligned}
\end{equation}
where
\begin{subequations}
\begin{eqnarray}
\hspace*{-3mm}W &:=& 1+ \left[\beta_{ab}\left(2\cos\theta - \beta_{ab}\right) + \beta_{a}^{2}\right]\csc^{2}\theta \,, \\
\hspace*{-3mm}\widehat{\Sigma} &:=& \Sigma  - \left(\beta^{2} + 2br\right) + \rg^{2} \, \beta_{b}\left(\beta_{b} - 2a_{\sub *} \cos\theta \right) \,, \\
\hspace*{-3mm}\widehat{\Delta} &:=& \Delta - \left(\beta^{2} + 2br\right) - \rg\left(\rg + 2b \right) \beta_{b}^{2} \,, \\
\hspace*{-3mm}\widehat{\mathcal{A}} &:=& \delta^{2} - a^{2} \widehat{\Delta} W^{2} \sin^{2}\theta \,, \\
\hspace*{-3mm}\delta &:=& r^{2}-2b\,r + a^{2} \,.
\end{eqnarray}
\end{subequations}
Here $b$ and $\beta$ denote the coupling parameters of the dilaton and axion fields, respectively, and have units of length.
For clarity, we have also defined:
\begin{equation}
\beta_{a} \equiv \frac{\beta_{\sub *}}{a_{\sub *}} \,, \qquad
\beta_{b} \equiv \frac{\beta_{\sub *}}{b_{\sub *}} \,, \qquad
\beta_{ab} \equiv \frac{\beta_{\sub *}}{a_{\sub *} b_{\sub *}} \,,
\label{EMDA_divergences}
\end{equation}
where $b_{\sub *}\equiv b/\rg$ and $\beta_{\sub *}\equiv \beta/\rg$ are dimensionless counterparts of these coupling parameters.
 
Inspecting eq.~\eqref{EMDA_divergences}, one immediately notices that for the effects of the axion field to be non-zero, the dilaton coupling and the black hole spin parameter must both be non-zero, i.e., one can obtain neither a spherically-symmetric nor an axisymmetric solely axion black hole solution.
By contrast, through setting the axion coupling to zero an axisymmetric dilaton black hole solution is obtained.
Furthermore, since the term $W$ is always multiplied by $a$, the spherically-symmetric EMDA black hole solution is recovered.
\subsection{The Johannsen-Psaltis metric}\label{JP_section}
The final metric we consider is the Johannsen-Psaltis (JP) metric, which is a general parametrized metric that describes the exterior solution of rapidly spinning black hole.
The JP metric introduces parametric deviations to the Kerr metric through adjustable ``deviation parameters''.
At lowest order, four such parameters exist, of which three are of physical relevance to this study.
The topological structure and geodesic integrability of the JP metric is well-studied in the literature, providing a reliable platform upon which to perform observational tests of astrophysical black holes (see~\citealt{Johannsen2013c,Medeiros2020}).
This enables the investigation of electromagnetic radiation produced in the vicinity of the event horizons of rapidly-spinning black holes which cannot be described by the Kerr solution, nor be admitted as a solution of the Einstein field equations of GR.
In Boyer-Lindquist-like coordinates, the JP line element \citep{Johannsen2013b} is written as:
\begin{equation}
\begin{aligned}
{\rm d}s^{2} = &-\frac{\widetilde{\Sigma} \, \mathcal{B}}{\mathcal{F}} \, c^{2}{\rm d}t^{2} - \frac{2 a \, \widetilde{\Sigma} \, \mathcal{C} \sin^{2} \theta}{\mathcal{F}} \, c \,{\rm d}t \, {\rm d}\phi \\
&+ \frac{\widetilde{\Sigma}}{\Delta A_{\sub 5}} \, {\rm d}r^{2} + \widetilde{\Sigma} \, {\rm d}\theta^{2} + \frac{\widetilde{\Sigma} \, \mathcal{D} \sin^{2}\theta}{\mathcal{F}} \, {\rm d}\phi^{2} \,,
\end{aligned}
\end{equation}
where:
\begin{subequations}
\begin{eqnarray}
\mathcal{B} &:=& \Delta - a^{2}A_{\sub 2}^{2}\sin^{2}\theta \,, \\
\mathcal{C} &:=& \left( r^{2} + a^{2} \right) A_{\sub 1} A_{\sub 2} - \Delta \,, \\
\mathcal{D} &:=& \left( r^{2} + a^{2} \right)^{2} A_{\sub 1}^{2} - a^{2} \Delta \sin^{2}\theta \,, \\
\mathcal{F} &:=& \left[ \left(r^{2} + a^{2}\right) A_{\sub 1} - a^{2} A_{\sub 2}\sin^{2} \theta \right]^{2} \,,
\end{eqnarray}
\end{subequations}
and the useful identity $\mathcal{F}\Delta\equiv 
\mathcal{B}\mathcal{D} + a^{2}\mathcal{C}^{2}\sin^{2}\theta$ holds.
Finally, the terms that depend on the Kerr metric deformation parameters are defined as follows
\begin{subequations}
\begin{eqnarray}
\widetilde{\Sigma} &:=& \Sigma + \rg^{2} \, \sum_{n=3}^{\infty} \epsilon_{\sub n} \, \left(\frac{\rg}{r}\right)^{n-2} \,, \label{tildesigma}\\
A_{\sub 1} &:=& 1 + \sum_{n=3}^{\infty} \alpha_{\sub 1 n} \left( \frac{\rg}{r} \right)^{n} \,, \\
A_{\sub 2} &:=& 1 + \sum_{n=2}^{\infty} \alpha_{\sub 2 n} \left( \frac{\rg}{r} \right)^{n} \,, \\
A_{\sub 5} &:=& 1 + \sum_{n=2}^{\infty} \alpha_{\sub 5 n} \left( \frac{\rg}{r} \right)^{n} \,.
\end{eqnarray}
\end{subequations}
The metric deformation parameters are all dimensionless and $\epsilon_{\sub 2}=\alpha_{\sub 12}=0$ in the above expressions.
As noted in \cite{Johannsen2013b}, this form of the metric satisfies asymptotic flatness, recovers the correct Newtonian limit, and satisfies current PPN constraints. At lowest order in the expansion, the metric depends only on $\alpha_{\sub 13}$, $\alpha_{\sub 22}$, $\alpha_{\sub 52}$, and $\epsilon_{\sub 3}$. When these parameters are set to zero the Kerr metric is recovered. In this study, we let $\alpha_{\sub 52}=0$, as it modifies only the $g_{rr}$ component of the metric and fixing it to zero ensures that the (outer) event horizon radius is always equal to that of the Kerr metric.

In choosing a parametrized metric, there are several desirable conditions for black hole spacetimes to be pathology-free, namely that they are: (i) stationary, (ii) axisymmetric, (iii) satisfy asymptotic flatness, and (iv) contain no singularities or closed timelike curves external to the outer event horizon.
To date, black hole solutions in alternative theories of gravity that satisfy these conditions, i.e., do not exhibit pathological behavior, generally admit a rank-2 Killing tensor and possess a fourth ``Carter-like'' integral of motion \citep{Carter1968},  \citep[see][for further information]{Vigeland2011}.
The existence of a rank-2 Killing tensor is a desirable property of black hole spacetimes, ensuring the geodesic motion is not chaotic. Whilst this restriction is not strictly necessary, deviations from geodesic separability, if present, are likely to be very small \citep[e.g.][]{Yagi2012} and have a negligible effect on black hole images.

The existence of a rank-2 Killing tensor ensures the geodesic motion
is not chaotic \cite[see, e.g.,][for an example of chaotic motion in the JP spacetime]{Zelenka2017}.
In this study we consider the subset of parametrized metrics which satisfy the
integrability condition, thereby ensuring the geodesic motion remains non-chaotic.
We note that this integrability condition is not strictly necessary and mapping to all known modified black hole solutions requires this condition be relaxed \citep[e.g.][]{Yagi2012}.
The effects of non-integrability on photon orbits have been investigated in recent studies \citep[see, e.g.,][]{Papas2018,Kostaros2022}.

Several different parametrized spacetime metrics exist in the literature \citep{Manko1992,Glampedakis2006,Vigeland2010,Vigeland2011,Johannsen2011,Rezzolla2014,Konoplya2016}, all differing in the extent to which the above four conditions are satisfied.
In subsequent works, some authors have extended some of these parametrized metrics to also ensure that the spacetimes yield either a separable Hamilton-Jacobi equation \citep[e.g., the JP metric employed in this study;][]{Johannsen2013} or separability in both the Hamilton-Jacobi \textit{and} Klein-Gordon equations \citep[e.g., the Konoplya-Rezzolla-Zhidenko metric;][]{Konoplya2018,Konoplya2021}.
Owing to the above considerations and the motivations at the start of Sec.~\ref{JP_section}, we choose the JP metric to model parametric deviations from astrophysical Kerr black holes in this study.

We note that the covariant plasma model we have developed in Paper~I and employed throughout the present work does not involve nor rely upon the existence of a fourth integral of motion.
The only aspect of our calculation that is affected by such an integral of motion is the ray tracing.
Several earlier studies have performed calculations of ray tracing in spacetimes that do not have a Carter-like constant.
They yielded conclusions regarding the size of the black-hole shadow, its relation to the size of the UPO, and its nearly circular shape that are identical to ours (see, e.g., \citealt{Johannsen2010a}, who used the quasi-Kerr spacetime by \citealt{Glampedakis2006}).

The only differences introduced to the shadow shape by the non-integrability of the spacetime are distinctive yet small deviations from the nearly circular shape in spacetimes with large deviations from Kerr \citep{Kostaros2022}, none of which alter the conclusions of this study.
\newline
\newline
\subsection{Characteristic radii}
Black hole spacetimes possess several different characteristic radii.  These radii define regions that characterize different physical properties of, and processes occurring in, a given spacetime geometry.
This study is concerned with three specific characteristic radii: the event horizon, $r_{\sub H}$, the unstable photon orbit (UPO), $r_{\sub UPO}$, and the innermost stable circular orbit (ISCO), $r_{\sub ISCO}$.

The covariant plasma model detailed in the next Section specifies a radial 4-velocity profile which is affected by the location of the ISCO. The radial velocity, in turn, determines the plasma number density and magnetic field strength.
The critical impact parameter of geodesics comprising a given black hole image, i.e., the apparent size of the black hole shadow boundary curve, is dependent on $r_{\sub UPO}$.
When calculating geodesic motion in arbitrary spacetime geometries, a-priori knowledge of the event horizon radius is necessary to specify an appropriate numerical cutoff radius, $r_{\sub cut}$, for numerical geodesic integration algorithms\footnote{Geodesics in axisymmetric non-Kerr spacetimes are usually numerically integrated in Boyer-Lindquist-like coordinates, i.e., coordinates where $g_{t\phi}$ is the only non-zero off-diagonal metric component. The event horizon is well known to be a removable co-ordinate singularity and the numerical integration of geodesics captured by the black hole in such coordinates must be terminated when sufficiently close to $r_{\sub H}$. In this study, we assume an inner cutoff radius $r_{\sub cut}=1.001~r_{\sub H}$, i.e., integration stops when geodesics are within $0.1\%$ of the event horizon.}. Owing to these considerations, in this study, it is necessary to determine these radii to very high precision.

In integrating the geodesic equations of motion, we typically employ a fourth-order Runge-Kutta-Fehlberg (RKF) algorithm with adaptive step-size control. In situations where the observer is placed much further from the black hole (fiducial distance of $\sim10^{4}~\rg$), e.g., in the large UPO cases considered later in this paper, we employ an eighth-order RKF method with adaptive step-size control. Throughout this study, we specify an integration tolerance of $10^{-12}$.

Appendix~\ref{Sec:GRRT_Code_Comparison} presents the results of a cross-validation between two independent relativistic radiative transfer codes, where we show that the leading discrepancy between the codes arises from the numerical accuracy by which the ISCO radius is determined. For this reason, we calculate all three critical radii to machine double precision, i.e., $\lesssim 10^{-16}$, prior to geodesic integration. We detail in Appendix~\ref{Appendix_Char_Radii}  the numerical procedure we use for determining the characteristic radii and present iso-contours of these radii for different coupling parameters of the EMDA and JP spacetimes.
\section{Plasma Model}
\label{sec:plasma}
\subsection{Simple plasma model}
In the subsequent exploration of black hole images and their dependence on the metric properties, we will first utilize a toy emissivity model where we allow the $1.3$~mm ($230$~GHz) emissivity to have a power-law dependence on the co-ordinate radius and an arbitrary scale height $h/r$:
\begin{equation}
    j(r,\theta)=j_{0} \, r^{-n}\exp\left\{-\frac{1}{2}\left[\frac{\theta-\pi/2}{(h/r)\pi/2}\right]^2\right\}\;.
\label{eq:emissivity}    
\end{equation}
This equation simplifies to a spherically symmetric emissivity as $h/r$ goes to infinity. This will allow us to separate the effects of the spacetime from those of the additional effects introduced by the plasma model and the relativistic Doppler shifts introduced by the motion of the gas. In this model, the plasma is considered to be in free-fall.
Explicit expressions for the four-velocities of free-falling particle geodesics in the EMDA and JP spacetimes are presented in Appendix~\ref{Appendix:free_fall}.

\subsection{Full covariant plasma model}\label{Sec:Full_Plasma_Model}
In the bulk of the results, we will employ an analytic model that is based on the solution of the conservation laws for mass, momentum, and energy that govern the dynamics and thermodynamics of the gas accreting towards the black hole. This covariant semi-analytic model is axisymmetric so that there is no dependence of any quantity on the azimuthal angle $\phi$. We review here the basic equations for completeness and refer the reader to \citetalias{Ozel2021} for the details and derivations. 

\subsubsection{Plasma model thermodynamic properties}
We set the equatorial electron density profile to
\begin{equation}
n_{\rm e ,eq}(\varpi) = \frac{\dot{M}}{4 \pi \sqrt{-g} \ (h/r) \ m_{\rm p} \left(-u^{r}\right)} \,, 
\label{eq:ne_3D}
\end{equation}
where $\varpi\equiv r\sin\theta$ is the equatorial radius, $m_{\rm p}$ is the proton mass (assuming a fully ionized hydrogen plasma), and $u^{r}$ is the $r-$component of the plasma four-velocity. The factor related to the determinant of the metric, $g=\det g_{\mu\nu}$, is evaluated at the same equatorial radius. We multiply this equatorial density profile $n_{\rm e, eq}$ with an exponential in the polar angle $\theta$, i.e., 
\begin{equation}
    n_{\rm e}(r,\theta)=n_{\rm e, eq}(\varpi)\exp\left\{-\frac{1}{2}\left[\frac{\theta-\pi/2}{(h/r)\pi/2}\right]^m\right\} \,, 
\label{eq:den_offeq}
\end{equation}
where the index $m$ determines the slope of the vertical density profile.
In the Newtonian limit and for an ion temperature that is constant with height, we find that $m=2$ and
\begin{equation}
\frac{h}{r} = \frac{1}{r u^\phi} \left(\frac{P}{\rho}\right)^{1/2} = \sqrt{(\hat{\gamma}-1)\zeta} \,. 
\label{eq:hr}
\end{equation}

The parameters $\hat{\gamma}$ and $\zeta$ arise from energy conservation arguments and will be introduced below.
We will use these expressions hereafter, unless specified otherwise. 

Solving the energy conservation equation leads to the following expression for the ion temperature:
\begin{equation}
    T_{\rm i}(r,\theta)= \frac{m_p c^2}{k_{\rm B}} \frac{R (\hat{\gamma}-1)}{(R+1)} {\cal V}\;, 
\label{eq:Ti_int}
\end{equation}
where $k_{\rm B}$ is the Boltzmann constant, $\hat{\gamma}$ is the effective adiabatic index of the plasma, $R$ is the ratio of the electron to ion temperature, and ${\cal V}$ is a density-weighted integral of the dissipation function for the process that is responsible for heating the flow.

Because the flow is radiatively inefficient, the ion temperature at any radius becomes comparable to the local virial temperature, evaluated appropriately for each spacetime. Following the procedure outlined in Appendix~A of \citetalias{Ozel2021} and keeping only the leading-order corrections introduced by the various metric deviation parameters, we write
\begin{equation}
    T_{\rm i}(r,\theta)= \frac{m_p c^2}{k_{\rm B}} \frac{R (\hat{\gamma}-1)}{(R+1)} \zeta \left(\frac{\rg}{r}\right)\mathcal{T}_{\sub c} \,, 
\label{eq:Ti_EMDA}
\end{equation}
where $\mathcal{T}_{\sub c}$ is a spacetime-dependent correction to the temperature which is equal to unity for the Kerr black hole. For the EMDA and JP metrics used in this study, these corrections are given respectively by 
\begin{equation}
\begin{aligned}
\mathcal{T}_{\sub c} = 1 &+ \frac{1}{3}\left(\frac{\rg}{r}\right) \Big[7b_{\sub *} + 4\left(b_{\sub *}+1\right)\beta_{b}^{2}\Big] \\
&+ \frac{1}{9}\left(\frac{\rg}{r}\right)^{2} \Big[b_{\sub *}\left(45 b_{\sub *} + 51.5\right) \\
&+ \left(b_{\sub *}+1\right)\left(42b_{\sub *}+11\right)\beta_{b}^{2}\Big] + \mathcal{O}\left(r^{-3}\right) \,,
\end{aligned}
\end{equation}
and
\begin{equation}
\mathcal{T}_{\sub c} = 1 +  \frac{1}{6}\left(\frac{\rg}{r}\right)^{2}\left( 10\alpha_{\sub 13} - \alpha_{\sub 52} - 5\epsilon_{\sub 3} \right) + \mathcal{O}\left(r^{-3}\right) \,.
\end{equation}
In both cases $\zeta$ is an order-unity factor. We then write the electron temperature as
\begin{equation}
T_{\rm e}(r,\theta)=\frac{T_i(r,\theta)}{R}\;.
\label{eq:Te}
\end{equation}
Finally, we specify the magnetic field everywhere such that the plasma-$\beta$ parameter is constant throughout the flow, i.e., such that
\begin{equation}
B(r,\theta) \propto \left[ n_{\rm e}(r,\theta) \ T_i (r,\theta)\right]^{1/2}\;.
\label{eq:B_beta}
\end{equation}
In the absence of synchrotron self absorption, which is negligible for the frequency and range of accretion rates of interest here, the overall normalization of the magnetic field can be specified at a fiducial equatorial location and scaled according to relation~(\ref{eq:B_beta}). We, therefore, write
\begin{equation}
    B\left(r,\theta\right)=B_{0} \left[\frac{ n_{\rm e}(r,\theta) \, T_{\rm i}\left(r,\theta\right) }{ n_{\rm e}\left(r_{0},\pi/2\right) \, T_{\rm i}\left(r_{0},\pi/2\right)}\right]^{1/2}\,,
    \label{eq:B0}
\end{equation}
such that $B_{0}$ is the strength of the magnetic field at the spherical radius $r_{0} \equiv r_{\sub ISCO} \left( a_{\sub *}{\rm =}0 ,\, \theta{\rm=}\pi/2\right)$, i.e., at the equatorial ISCO radius for a non-spinning black hole in the spacetime being considered.
This particular normalization for non-Kerr spacetimes is chosen in order to smoothly recover the Kerr expression in \citetalias{Ozel2021}, while properly accounting for cases of large metric deviation parameters where the ISCO radius for $a_{\sub *}=0$ can deviate significantly from the $6~\rg$ Schwarzschild value.

We use the analytic fitting formula for the angle-averaged emissivity derived by~\citet{Mahadevan1996}, which is accurate to within 2.6\% for all temperatures and frequencies of interest:
\begin{equation}
 \label{eq:jtot}
 j_\nu = \frac{n_{\rm{e}} \, e^2 \nu}{\sqrt{3} \, c \, K_2(1/\Theta_e)} \, {\rm M}(x_{\rm M}) \,,
\end{equation}
with ${\rm M}(x_{\rm M})$ given by:
\begin{equation}
\begin{aligned}
{\rm M}(x_{\rm M}) = &\frac{4.0505\,\mathpzc{a}}{x_{M}^{1/6}} \left(1 + \frac{0.40\,\mathpzc{b}}{x_{\rm M}^{1/4}} +  \frac{0.5316\,\mathpzc{c}}{x_{\rm M}^{1/2}}\right) \\
&\times \exp\left(-1.8896 \; x_{\rm M}^{1/3}\right)\,.
\label{eq:mxm}
\end{aligned}
\end{equation}
Here, $\nu_{b} \equiv e B / \left(2 \pi m_{\rm e} c\right)$ is the cyclotron frequency and
\begin{equation}
 \label{eq:xm}
 x_{\rm M} \equiv \frac{2 \nu}{3 \, \nu_{b} \, \Theta_{\rm e}^2}\;,
\end{equation}
where $\Theta_{\rm e} \equiv k_{\rm B} T_{\rm e} / \left(m_{\rm e} c^2\right)$ is the dimensionless electron temperature, $e$ is the electron charge, $m_{\rm e}$ is the electron mass, and $K_2(x)$ is the modified Bessel function of the second kind of order two and with argument $x$.
The best fit values of the coefficients $\mathpzc{a},\,\mathpzc{b}$, and $\mathpzc{c}$ for different temperatures are given in \citet{Mahadevan1996}.
\subsubsection{Plasma model four-velocity profile}\label{Sec:full_plasma_4_vel}
For each spacetime, the equatorial radius of the ISCO provides a natural separatrix to specify the plasma velocities.
On the equatorial plane and outside the ISCO, we set the azimuthal component of the 4-velocity, $u^\phi$, equal to the orbital velocity of test particles at the same location.
The radial velocity profile depends on the efficiency of the angular momentum transport by material and magnetic stresses.
In order to allow for a general form which does not depend on the specifics of angular momentum transport, we write the radial velocity as:
\begin{equation}
    u^r_{\rm eq}(r)=-\eta \left(\frac{r}{r_{\sub ISCO}}\right)^{-n_{r}} \,,
    \label{eq:vr}
\end{equation}
where $\eta$ and $n_{r}$ are free parameters. 

On the equatorial plane and inside the ISCO, we calculate the azimuthal and radial velocity profiles by following the trajectories of test particles that free-fall with the energy and angular momentum of the plasma at the ISCO radius. Throughout the flow, we assume that the polar component of the 4-velocity is zero, i.e., $u^{\theta}=0$, and calculate $u^{t}$ by imposing the condition $g_{\mu\nu}u^{\mu}u^{\nu}=-1$. 

In order to model the plasma velocities off the equatorial plane, we use the fact borne out from semi-analytic models and GRMHD simulations that the azimuthal components $u^{\phi}$ are approximately constant on spherical surfaces and that the radial components $u^{r}$ are approximately constant on cylindrical surfaces. We again refer the reader to \citetalias{Ozel2021} for the details of the model.

We choose a set of fiducial values for the plasma parameters to employ for the majority of the paper that are representative of disk accretion and are also consistent with the numerical solutions obtained in GRMHD models of M87. In particular, unless stated otherwise, we use $\eta=0.1$ and $n_r=1.5$ for the radial velocity profile (eq.~[\ref{eq:vr}]), $\zeta=0.25$ and $\hat{\gamma}=5/3$ for the ion temperature, $R=5$ for the electron temperature (eq.~[\ref{eq:Te}]), and $B_{0}=20$~G for the magnetic field scale (eq.~[\ref{eq:B0}]). These values lead to $h/r\simeq0.4$ for the disk scale height (eq.~[\ref{eq:hr}]).
Although we have demonstrated the lack of sensitivity of our conclusions on the specific plasma parameters in \citetalias{Ozel2021}, we nevertheless use a second set of plasma parameters to show that this lack of sensitivity is not specific to the Kerr metric.
\begin{figure*}[htb!]
\begin{center}
\includegraphics[width=0.49\textwidth]{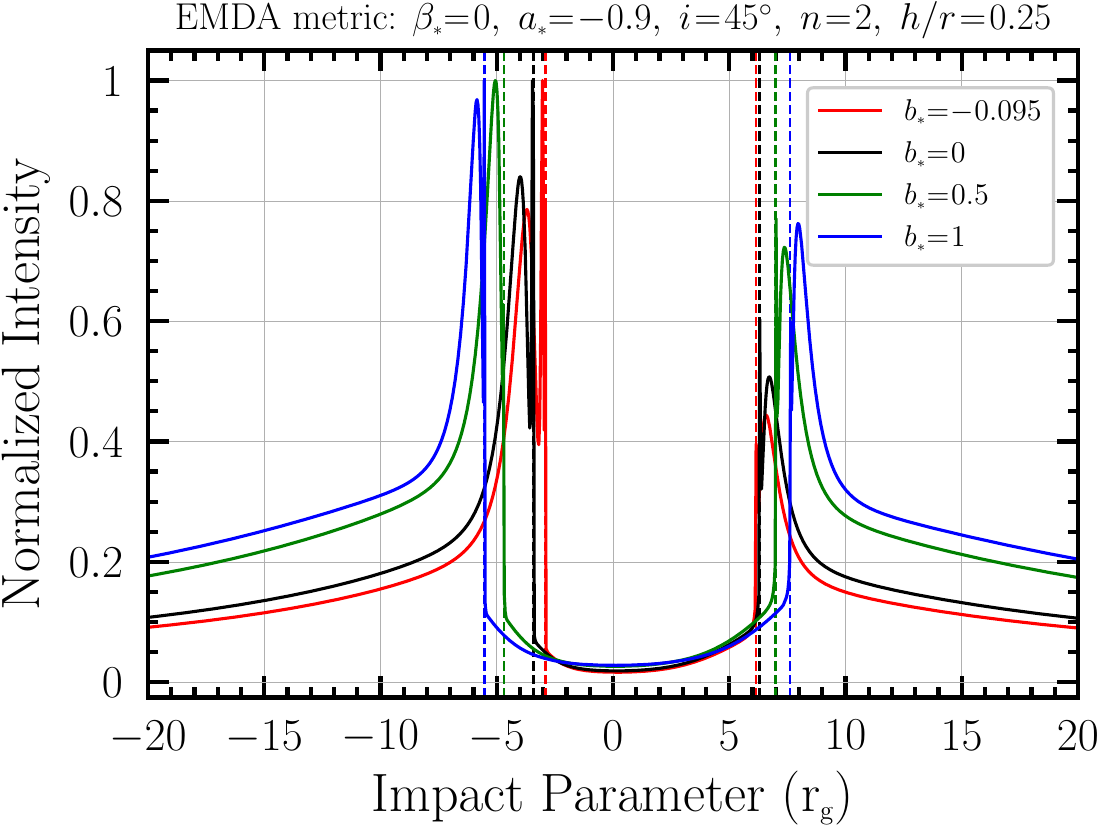}\hfill
\includegraphics[width=0.49\textwidth]{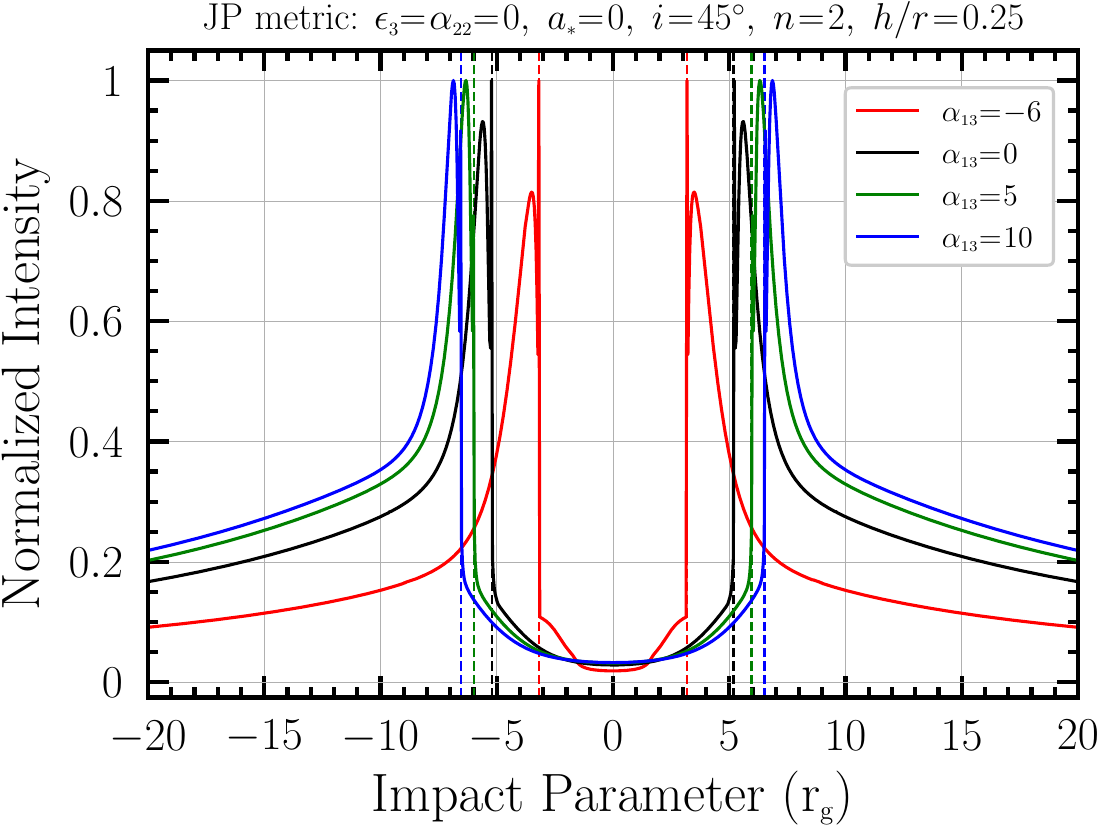}
\end{center}\vspace*{-3mm}
\caption{\footnotesize The effect of varying spacetime geometry on horizontal image cross-sections produced by free-falling matter with a simple power-law emissivity profile given by eq.~\eqref{eq:emissivity}.
The panels present black holes described by the EMDA metric with a large retrograde spin and varying dilaton field coupling (left) and the JP metric with zero spin and varying $\alpha_{\sub 13}$ parameter (right).
Metric parameter values of $b_{\sub *}$ and $\alpha_{\sub 13}$ are chosen to sample from the minimum allowed value to considerably larger than the reference Kerr value (black curve in both panels).
Vertical dashed colored lines delineate the left and right critical impact parameters of each cross-section.
}
\label{fig:Analytic_Freefall}
\end{figure*}
\subsection{Ray-Tracing and General-Relativistic Radiative Transfer}
The geodesic equations of motion and the equations of general-relativistic radiative transfer (GRRT) are solved using the \texttt{BHOSS} code \citep{Younsi2012,Younsi2016,Younsi2020}.
In this study the equations of motion governing null geodesics (hereafter rays) are integrated as:
\begin{equation}
\frac{{\rm d} k^{\mu}}{\mathrm{d}\lambda} = -\Gamma^{\mu}_{\phantom{\mu}\alpha\beta} \, k^{\alpha} k^{\beta} \,, \qquad \frac{{\rm d} x^{\mu}}{\mathrm{d}\lambda} = k^{\mu} \,,
\label{eqn:GEO}
\end{equation}
where $\lambda$ is the affine parameter parametrizing the ray, $k^{\mu}$ is the ray's four-momentum, and $x^{\mu}$ is its position four-vector. After specifying the expressions for the covariant metric tensor components, $g_{\mu\nu}$, the Christoffel symbols, $\Gamma^{\mu}_{\phantom{\mu}\alpha\beta}$, are computed via centered finite differences, with the contravariant metric tensor computed via LU-decomposition of $g_{\mu\nu}$. Equations~\eqref{eqn:GEO} are typically integrated using a fourth-order Runge-Kutta-Fehlberg (RKF) method with adaptive step-size control, and in instances where the spacetime under consideration strongly deviates from Kerr and the required precision is necessarily higher, a sixth-order RKF method is used.
In numerically integrating the black hole shadow boundary curves using a bisection method (see \citealt{Younsi2016}), an eighth-order RKF method is used.

After determining the ray trajectories, we solve the GRRT equations along these rays, accounting for emission and the effects of attenuation of the ray intensities by the intervening media between the black hole event horizon and the observer. The equation expressing the covariant radiative transport of unpolarised, unscattered radiation \citep[see, e.g.,][]{Lindquist1966,Fuerst2004,Younsi2012} is written as:
\begin{equation}
\frac{{\rm d}\mathcal{I}_{\nu}}{{\rm d}\lambda} = -k^{\alpha} u_{\alpha}\left( -\chi_{\nu,0} \, \mathcal{I}_{\nu}+\frac{j_{\nu,0}}{{\nu_{0}}^3} \right) \,,
\end{equation}
where $u^{\alpha}$ is the plasma four-velocity, $\nu$ denotes the observing frequency, subscript ``0'' denotes quantities evaluated in the co-moving frame of the accretion flow, $\chi_{\nu}$ and $j_{\nu}$ denote the frequency-dependent absorption and emission coefficients, respectively, and $\mathcal{I}_{\nu} \equiv I_{\nu}/\nu^{3}$ is the (frequency-dependent) Lorentz-invariant intensity and $I_{\nu}$ the corresponding specific intensity.

We solve these equations numerically in decoupled form, as originally expressed in \cite{Younsi2012}, which for an observer (subscript ``obs'') at $\lambda_{\rm obs}$ may be written as:
\begin{subequations}
\begin{eqnarray}
\frac{\mathrm{d} \tau_{\nu}}{\mathrm{d} \lambda} &=& g^{-1} \chi_{\nu,0} \,, \label{eq:tau_GRRT} \\
\frac{\mathrm{d} \mathcal{I}_{\nu}}{\mathrm{d} \lambda} &=& g^{-1} \left(\frac{j_{\nu,0}}{{\nu_{0}}^{3}}\right) \mathrm{e}^{-\tau_{\nu}} \,, \label{eq:I_GRRT}
\end{eqnarray}
\end{subequations}
where $g\equiv \nu/\nu_{0}=(k^{\beta}u_{\beta}|_{\lambda_{\rm obs}})/(k^{\alpha}u_{\alpha}|_{\lambda})$ is the relative energy shift of the photon between the observer and comoving frames, and $\tau_{\nu}$ denotes the frequency-dependent optical depth.
We solve these equations using a simple first-order Euler method, with the step size determined from the RKF geodesic integration.
This formulation has the advantage of integrating the GRRT equation in tandem with the geodesic equations, i.e., from observer to source.
This avoids storing geodesics in memory for the GRRT integration, enables the tracking and truncation of ray optical depth, and is computationally fast and extendable to multi-frequency integration.
In this study, we assume that the emission is optically thin, i.e., $\chi_{\nu}=0$, and therefore integrate eq.~\eqref{eq:I_GRRT} alone.
\begin{figure*}[htb!]
\begin{center}
\includegraphics[width=0.99\textwidth]{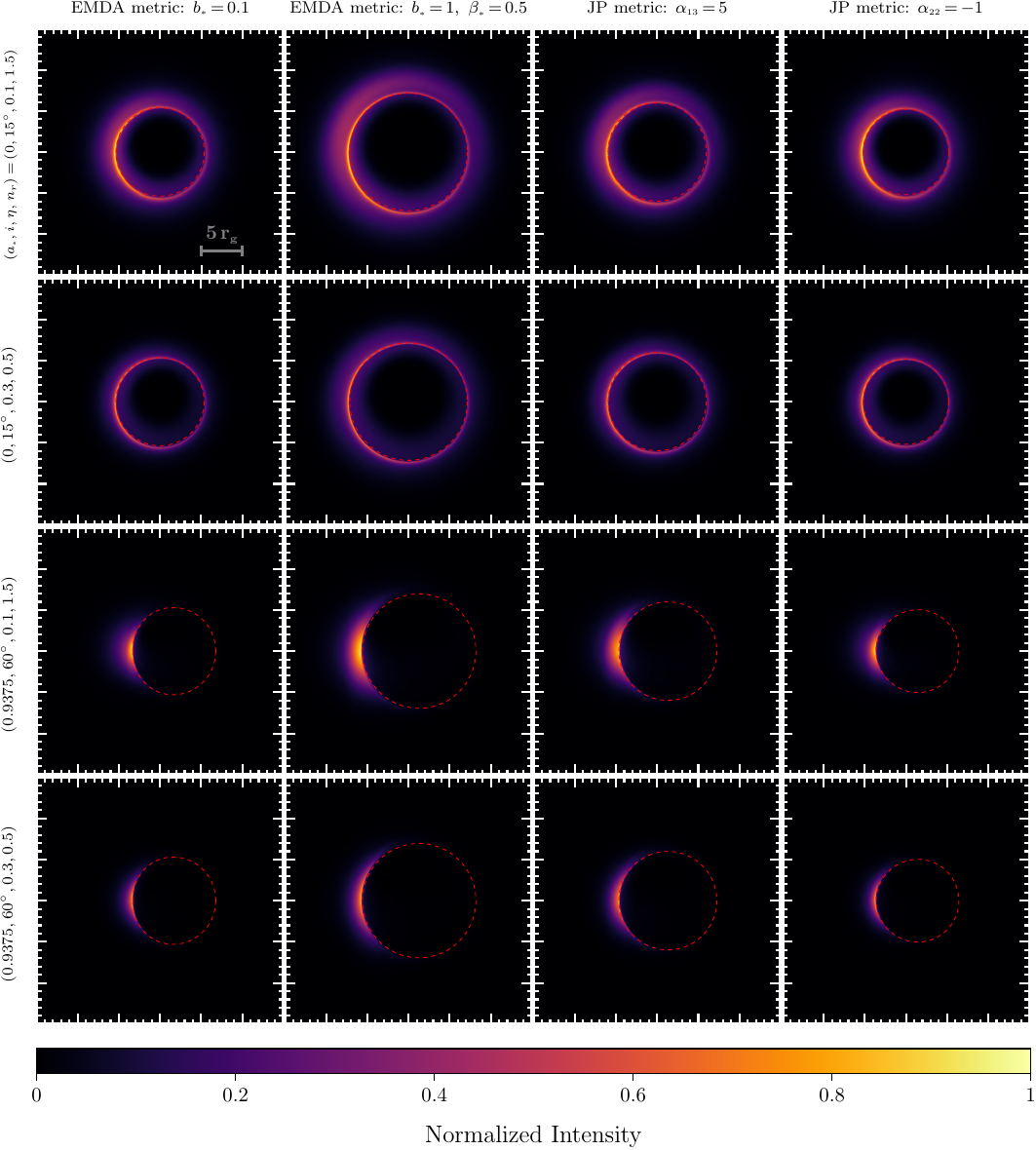}
\end{center}\vspace*{3mm}
\caption{\footnotesize 1.3 mm images from the full covariant plasma model for the EMDA and JP spacetimes. Image panels are individually normalized such that the brightest pixel intensity is unity.
The field of view is $[-15~\rg,\, 15~\rg]$ in both directions.
From left to right, the four columns correspond to: (i) the EMDA metric with $b_{\sub *}=0.1$, (ii) the EMDA metric with $b_{\sub *}=1,\,\beta_{\sub *}=0.5$, (iii) the JP metric with $\alpha_{\sub 13}=5$, and (iv) the JP metric with $\alpha_{22}=-1$.
Unless stated, all other metric parameters are set to zero.
The four rows present different values of $(a_{\sub *},i,\eta,n_{r})$.
Images in the top two rows fix $a_{\sub *}=0$ and $i=15^{\circ}$, varying $(\eta,n_{r})$, i.e., they alter the radial 4-velocity profile in eq.~\eqref{eq:vr}.
The bottom two rows present images with fixed $a_{\sub *}=0.9375$ and $i=60^{\circ}$, again varying $(\eta,n_{r})$ as in the upper two rows.
The dashed red curve in each panel marks the black hole shadow boundary.
}
\label{fig:Full_Plasma_Model}
\end{figure*}
\section{Results}\label{sec:results}
Having specified the metric and plasma models used in this paper in the previous two sections, we turn to exploring the properties of black hole images that result from these models. We separately vary the spacetime and plasma parameters in order to disentangle their effects. As in \citetalias{Ozel2021}, we begin by first utilizing a simple power-law emissivity model before focusing on the convolution of effects arising from both the full covariant plasma model and the spacetime geometry.

\subsection{Simple power-law emissivity with free-fall plasma}\label{Sec:simple}

We calculate images using the power-law emissivity profile of eq.~\eqref{eq:emissivity} for a power-law index $n=2$, a disk scale height $h/r=0.25$, and an observer inclination of 45$^\circ$.
We also assume that the plasma is free-falling (see Appendix~\ref{Appendix:free_fall} for the derivation of the relevant 4-velocities). Figure~\ref{fig:Analytic_Freefall} shows horizontal image cross-sections for different parameters of the EMDA and JP metrics and for different values of the black hole spin. We show the locations of the critical impact parameters as vertical dashed lines. 

Comparing these cross-sections to those in Figure~1 of \citetalias{Ozel2021} for the Kerr metric, we see that all of the same universal features persist for images in the non-Kerr case.
In particular, the peak of the emission always occurs very close to the critical impact parameter\footnote{Formally, the intensity approaches infinity at the critical impact parameter. As this narrow feature is indistinguishable from the broader ring at the current EHT resolution, we indicate the broad ring when we refer to peak intensity.} even when the latter varies substantially with the metric parameters.
Inside the critical impact parameter, there is always a sharp brightness depression that would be identified with the black hole shadow.
Finally, in the effective absence of azimuthal velocities in the free-falling regime, brightness asymmetries in the image are caused almost exclusively by frame-dragging effects.
These asymmetries are relatively minor unless frame dragging is enhanced substantially beyond the Kerr value (e.g., by increasing the $\alpha_{\sub 22}$ parameter in the JP metric to significantly large values). 

\begin{figure*}[htb!]
\begin{center}
\includegraphics[width=0.98\textwidth]{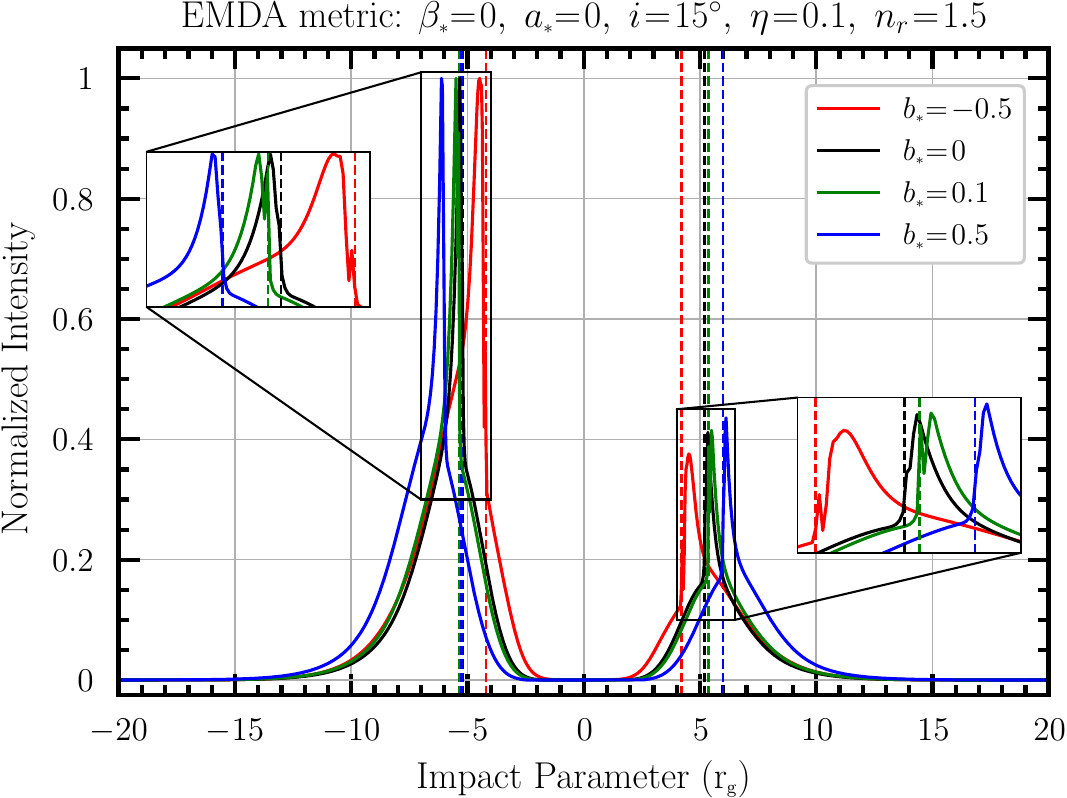}
\end{center}\vspace*{-3mm}
\caption{\footnotesize Horizontal normalized intensity cross-sections of $1.3$~mm non-rotating EMDA black-hole images calculated from the full covariant plasma model.
The axion field coupling parameter is set to zero and the observer inclination angle is $15^{\circ}$.
Solid red ($b_{\sub *}=-0.5$), black ($b_{\sub *}=0$, i.e., Schwarzschild), green ($b_{\sub *}=0.1$), and blue ($b_{\sub *}=0.5$) curves show the four different horizontal intensity cross-sections.
Vertical dashed lines represent the critical impact parameters for the four cases.
Left and right zoomed-in regions show the behavior of the four image intensity cross-sections of each image in the vicinity of its critical impact parameters.
The zoomed-in regions demonstrate that a final local maximum in intensity is always observed at, or in close proximity but external to, the left and right critical impact parameters of each image cross-section.
}
\label{fig:X-section_EMDA}
\end{figure*}
\begin{figure}[htb!]
\begin{center}
\includegraphics[width=0.47\textwidth]{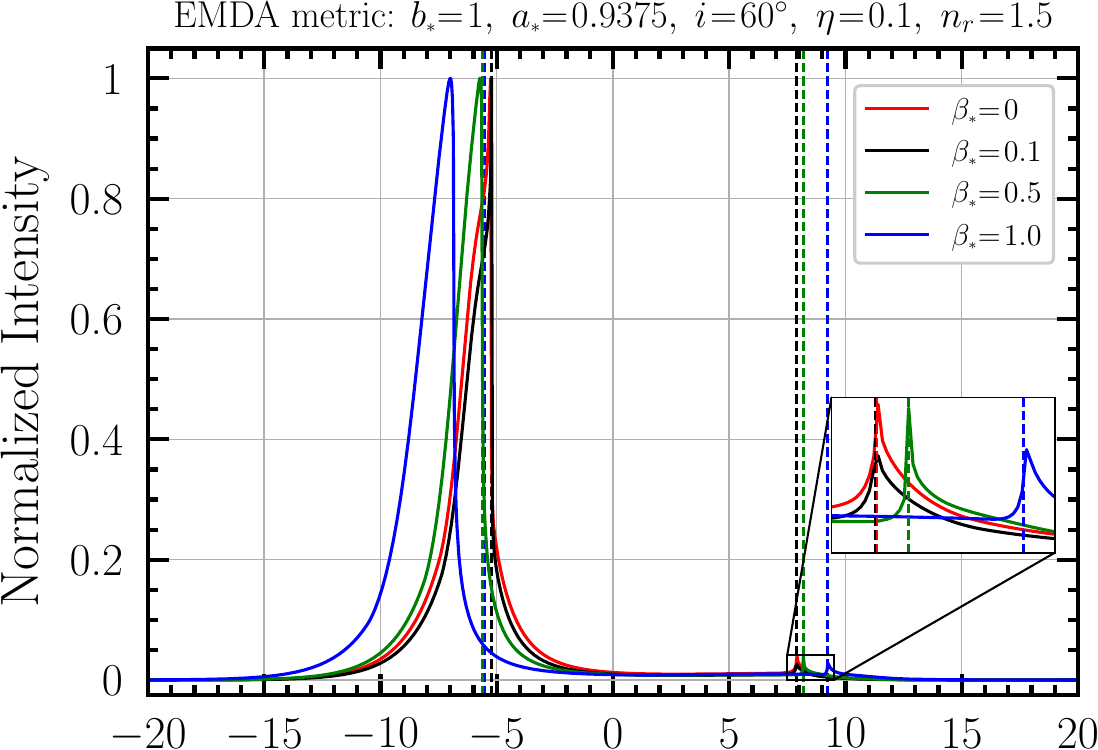}\\
\vspace*{4.0mm}
\includegraphics[width=0.47\textwidth]{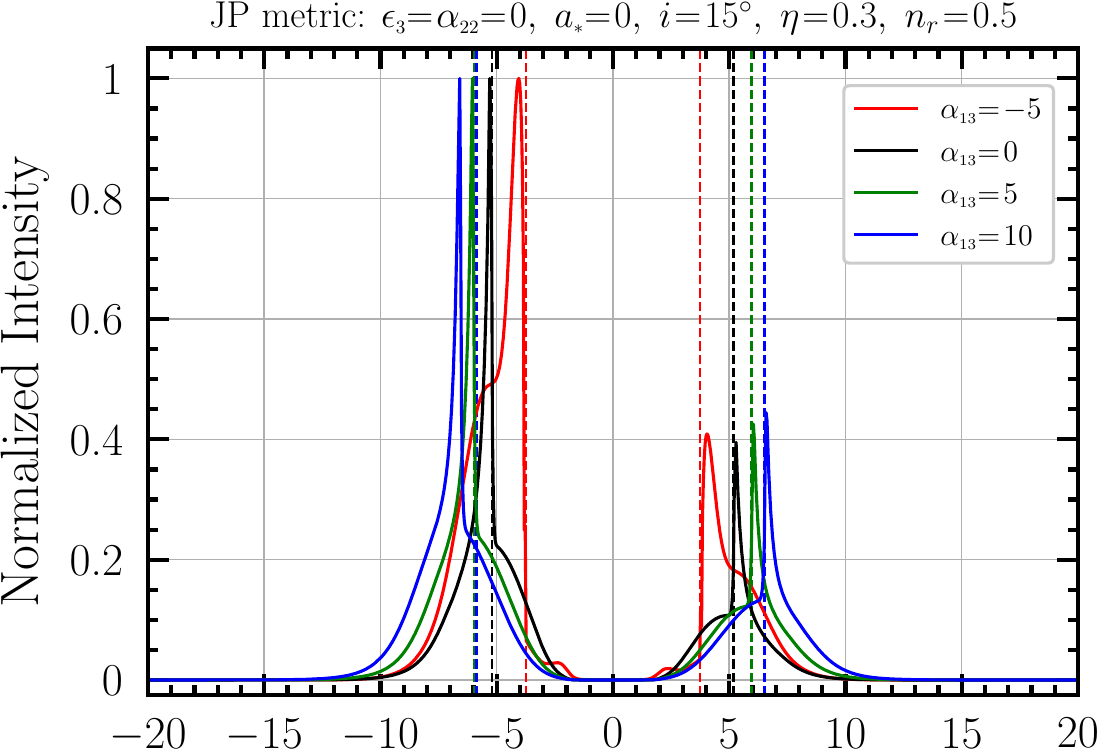}\\
\vspace*{4.0mm}
\includegraphics[width=0.47\textwidth]{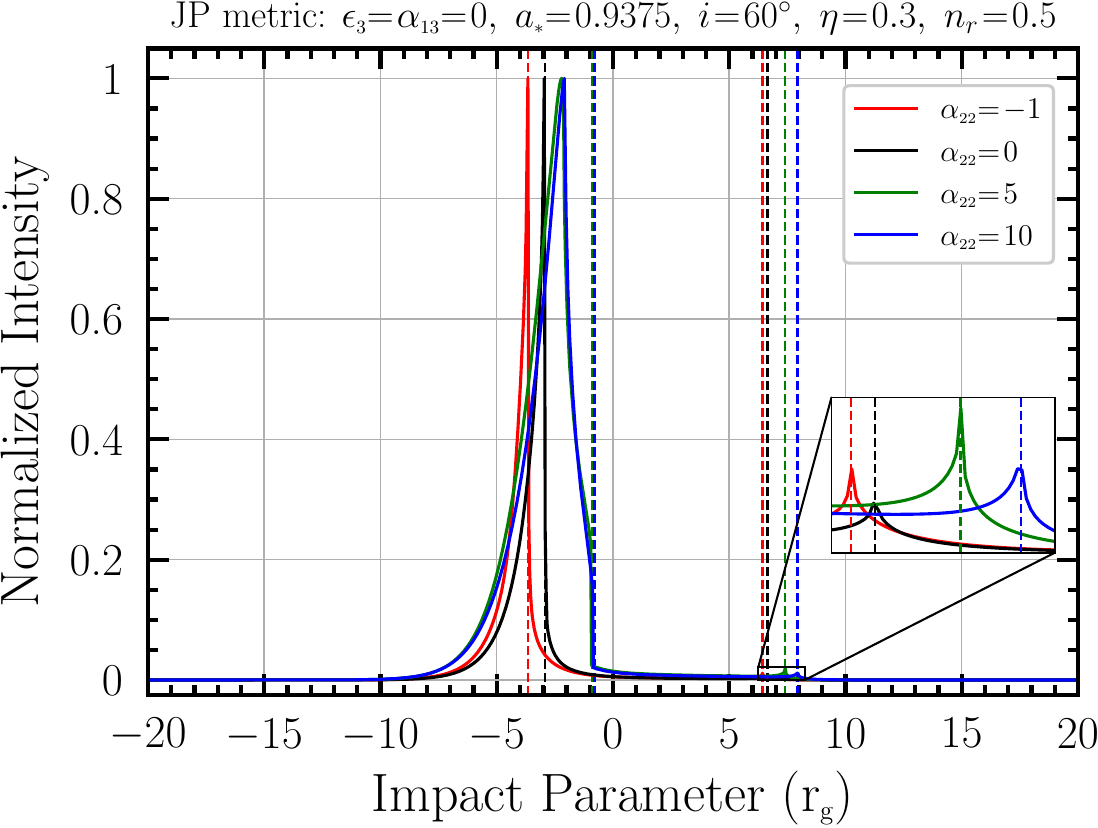}
\end{center}\vspace*{-3mm}
\caption{\footnotesize Horizontal normalized intensity cross-sections of EMDA and JP black hole images from the full covariant plasma model, at $1.3$~mm.
Panels represent: (top) a rapidly-spinning EMDA black hole, with a fixed dilaton field coupling and varying axion field coupling, (middle) a non-rotating JP metric with varying $\alpha_{\sub 13}$ deviation parameter, and (bottom) a rapidly-spinning JP metric with varying $\alpha_{\sub 22}$ parameter. Main plots and zoomed-in regions again demonstrate the near-coincidence of the peak brightness (albeit externally) with the critical impact parameter.
}
\label{fig:X-sections_EMDA_JP}
\end{figure}
\subsection{Full plasma model}
For the remainder of this paper, we employ the full covariant plasma model to describe the accreting material and its emission characteristics around the black hole (see Sec.~\ref{Sec:Full_Plasma_Model}).
In addition, we incorporate realistic velocities, which include substantial azimuthal components that can give rise to images with a large brightness asymmetry.

Figure~\ref{fig:Full_Plasma_Model} shows a selection of model images with the full covariant plasma model, for different values of the EMDA and JP metric parameters as well as for different black-hole spins, observer inclination angles, and plasma parameters. In all cases, the images of these optically-thin accretion flows are ring-like or crescent-like, with a deep central brightness depression, despite the significantly different spacetime geometries being explored. The main effect of changing the plasma parameters (cf.\ first and second rows in the figure) is to alter the width of the emission ring: shallow density profiles lead to broader rings and vice-versa. Increasing the observer inclination (cf.~the top two and bottom two rows) increases the brightness asymmetry between the approaching and receding part of the image, which is caused by relativistic Doppler effects arising from the azimuthal velocity component of the accreting material. For the model parameters displayed here, low inclinations correspond to ring-like images, whereas high inclinations correspond to crescents (see \citealt{Medeiros2021} for exceptions to this behavior).

The diameters of the bright image rings (or crescents) do not change appreciably when the black-hole metric is fixed, i.e., moving down any column in this figure. On the other hand, changing the metric or its parameters, i.e., comparing different columns in the figure, alters the diameters of the bright rings. However, in all cases the diameters of the images scale with the diameters of the shadows, which are shown as dashed red lines. As a result, measuring the diameter of the ring for a black hole of known mass can indeed be used as a test of the metric \citep{Psaltis2020}. 

In Figure~\ref{fig:X-section_EMDA}, we explore in more detail the image brightness near the critical impact parameters for a number of EMDA metrics with no spin. As in the case of the analytic emissivity models discussed in Sec.~\ref{Sec:simple}, strong gravitational lensing near the unstable photon orbits causes a sharp increase in the brightness at impact parameters close to the critical values (denoted in the figure by vertical dashed lines). Even when the locations of the critical impact parameters change as the metric properties are varied, the dominant brightness peaks are displaced concurrently.

Finally, we present in Figure~\ref{fig:X-sections_EMDA_JP} image brightness cross-sections of EMDA and JP metrics with different deviation parameters, black-hole spins, observer inclination angles, and plasma model parameters. 
This broader exploration confirms that:
\begin{enumerate}[label=(\roman*),noitemsep]
\item the diameter of the bright ring follows the shadow diameter closely,
\item it is the critical impact parameter that plays the dominant role in determining the properties of the image,
\item images from optically-thin accretion flows remain narrow even in non-Kerr spacetimes.
\end{enumerate}

In the following subsections we explore a comprehensive library of non-GR black-hole images, considering the variation of ten different model parameters and subsequently quantifying the relationship between image properties and spacetime metric parameters.
\newline
\newline
\subsection{Image library and \texorpdfstring{$\alpha$}{TEXT}-calibration}
\label{SSec:Library_alpha}
The close proximity of the peak brightness of the black-hole image to the critical impact parameter makes it possible to use the characteristics of black hole images to perform measurements of the underlying spacetime properties and conduct tests of the Kerr metric.
Accomplishing this in a quantitative manner requires establishing a relationship, together with its corresponding uncertainty, between the diameter $d_{\rm im}$ of the peak brightness, which is the observed quantity, and the diameter $d_{\rm sh}$ of the black-hole shadow, which probes the metric.
To this end, we introduced in previous work the calibration parameter:
\begin{equation}
    \alpha_1 \equiv \frac{d_{\rm im}}{d_{\rm sh}} \,,
\end{equation}
and used numerical and analytic models within GR to measure the range of possible values of this parameter in the Kerr spacetime (see \citetalias{Ozel2021})\footnote{Calibration of the image diameter to the shadow diameter also has a component that arises from model fitting and imaging algorithms.
To distinguish this component from the purely theoretical displacement explored here, we refer to the latter as $\alpha_{1}$.}.
Here, we extend this calculation to images in non-Kerr spacetimes. 

For the majority of the spacetime parameters explored in this study, the shadow is nearly circular and its diameter can be characterized by a single number for present purposes. Using the numerical method discussed earlier, we calculate the diameter of the shadow along different azimuthal cross-sections and adopt the average value as $d_{\rm sh}$.
We then measure the image diameter $d_{\rm im}$ using a characterization algorithm, as before, where we filter the image at the resolution of the EHT array, calculate analytically the center of each black hole shadow, and measure over a number of azimuthal cross-sections the distance of the peak emission from that center (see \citetalias{Ozel2021} for additional details).
We then identify the the diameter of the bright emission ring as being twice the value of the median distance. Within the same algorithm, we also measure the full width at half maximum (FWHM) of the shadow ring by approximating the brightness distribution along each azimuthal cross-section with an asymmetric Gaussian and then calculate the median value of the widths of these Gaussians.

\begin{figure*}[htb!]
\begin{center}
\includegraphics[width=0.49\textwidth]{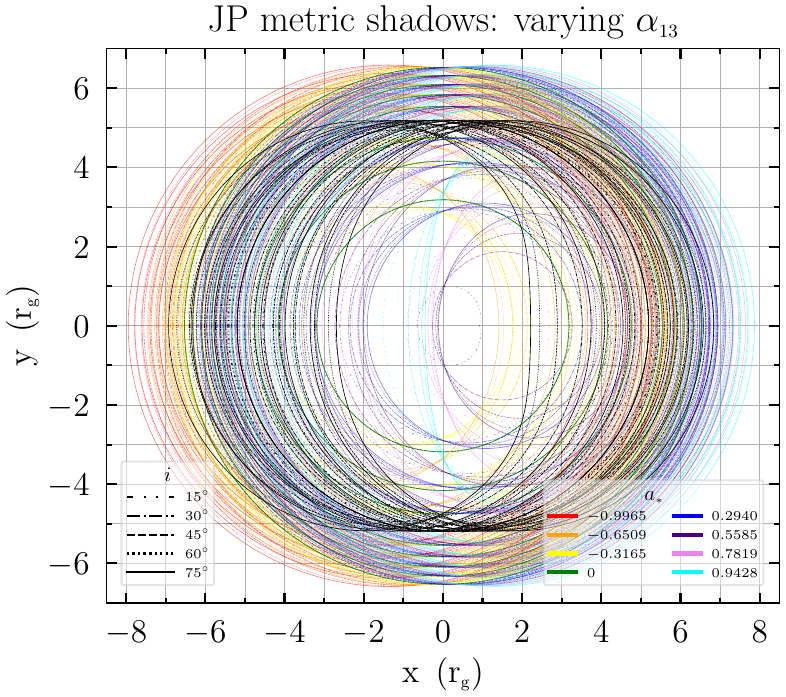}\hfill
\includegraphics[width=0.49\textwidth]{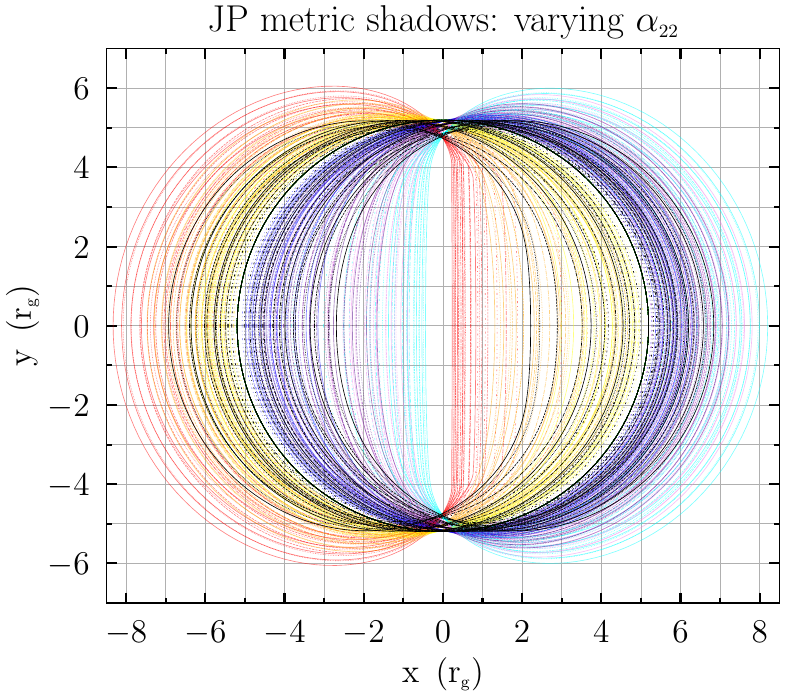}
\end{center}
\vspace*{-3mm}
\caption{\footnotesize Shadow boundary curves of black holes in the JP image library (see Table~\ref{Tab:JP_library_params}).
Coloured lines denote different dimensionless spin parameters and their line styles correspond to different observer inclination angles (see legends in left panel).
Kerr shadows are shown as black curves.
Left and right panels present shadows cast by non-Kerr black holes when varying $\alpha_{\sub 13}\in [-6,\,10]$ and $\alpha_{\sub 22}\in [-2,\,10]$, respectively, whilst fixing all other deviation parameters to zero.
For $\alpha_{\sub 13}<0$, shadows are always interior to their corresponding (black) Kerr shadows (left panel).
Most notably, for the extreme case of $\alpha_{\sub 13}=-6, \, a_{\sub *}=0.5585,\, i=15^{\circ}$, the shadow drastically decreases in angular size and is contained within the region $\sim[-1,\,1]\times[-1,\,1]$ (in units of $\rg$).
Conversely, when $\alpha_{13}$ is increased to larger positive values the shadows substantially increase in angular size as compared to their Kerr counterparts and become more circular, masking the effects of spin-induced (and observer inclination-induced) asymmetry.
By contrast, varying $\alpha_{\sub 22}$ leads to pronounced deviations from Kerr with significantly enhanced asymmetry (right panel).
This can be seen when comparing the red and cyan curves ($a_{\sub *}=-0.9965$ and $a_{\sub *}=0.9428$) with respect to the horizontal critical impact parameters of the Kerr shadows.
Note that the parameter $\epsilon_{\sub 3}$ alters the location of $r_{\sub ISCO}$ while leaving $r_{\sub UPO}$ unchanged, hence the shadow boundary curve properties are independent of $\epsilon_{\sub 3}$.
}
\label{fig:JP_shadows}
\end{figure*}

We carry out these measurements for a wide range of $1.3$~mm images obtained using the spacetimes and plasma models described in Sections~\ref{sec:metrics} and \ref{sec:plasma}. 

For the EMDA and JP metrics we consider a fine-grained sampling of model parameters: $(b_{\sub *},\, \beta_{\sub *})$ for the EMDA metric, $(\epsilon_{\sub 3}\,, \alpha_{\sub 13},\, \alpha_{\sub 22})$ for the JP metric, and $(i,\, a_{\sub *},\, \eta,\, n_{r}\,, R,\, B_{0},\, \zeta)$ for both metrics, all of which are shown in Table \ref{Tab:JP_library_params}, yielding a library of $\sim 2\times 10^{5}$ images.
Each parameter affects a different aspect of the image and the spacetime.
Varying $a_{\sub *}$ controls the location of the characteristic radii of the spacetime and changes physical effects such as the degree of frame-dragging, and, in the case of deviations from Kerr, the multipolar structure of the black hole.
The EMDA field couplings $b_{\sub *}$ and $\beta_{\sub *}$ alter the spacetime geometry and are explored in Figures~\ref{fig:JP_ISCO}--\ref{fig:EMDA_ISCO}.
Deviation parameters $(\epsilon_{\sub 3},\,\alpha_{\sub 13},\,\alpha_{\sub 22})$, which are varied one at a time while the others are set to zero, control the strength of deviation of the spacetime from the Kerr metric. For instance, as the value of $\alpha_{\sub 22}$ increases, the quadrupole moment of the black hole and the effects of frame-dragging are significantly enhanced beyond what is possible for even an extremal Kerr black hole. This leads to highly asymmetric shadow boundary curves as well as UPO radii very close to $r_{\sub H}$ (see Appendix \ref{Sec:Appendix_Library} for an example of this).

Adjusting $(\eta,n_{r})$ directly alters the plasma radial 4-velocity and it profile steepness as a function of radius, affecting the amount of extended emission produced in black hole images within the ISCO radius in particular, but also in its outer vicinity. The parameters $R$ and $B_{0}$ control the radiative properties of the plasma. As noted in \citetalias{Ozel2021}, we do not consider values of $R=1$ since this is inconsistent with the assumption of a radiatively-inefficient accretion flow, as anticipated for EHT target sources (see, e.g., \citealt{PaperV}). We also avoid consideration of models with $R>10$, because larger values of $R$ strongly suppress emission from  the accretion flow in the models. We choose values of $B_{0}$ which, based on isothermal one-zone modeling, yield electron temperatures consistent with the observed maximum brightness temperature of M87 \citep{EHTC2021_PaperVIII}. The largest value of $B_{0}=50~\mathrm{G}$, which is slightly larger than the upper-bound of the one-zone model, allows us to probe the broader range of potential electron number densities possible within non-Kerr accretion flow models.
Finally, in all models, we fix the adiabatic index as $\hat{\gamma}=5/3$ while varying $\zeta$, in effect adjusting the disk scale height.
We find that library image morphologies exhibit the weakest dependence on $\zeta$.
\begin{figure*}[htb!]
\vspace*{7mm}
\begin{center}
\includegraphics[width=0.5\textwidth]{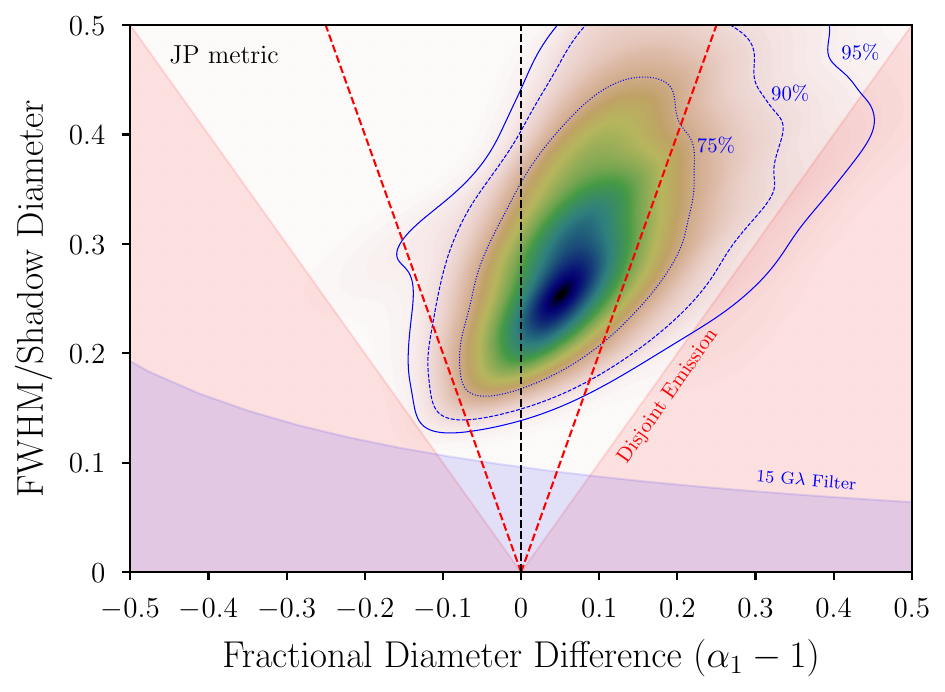}\hfill
\includegraphics[width=0.5\textwidth]{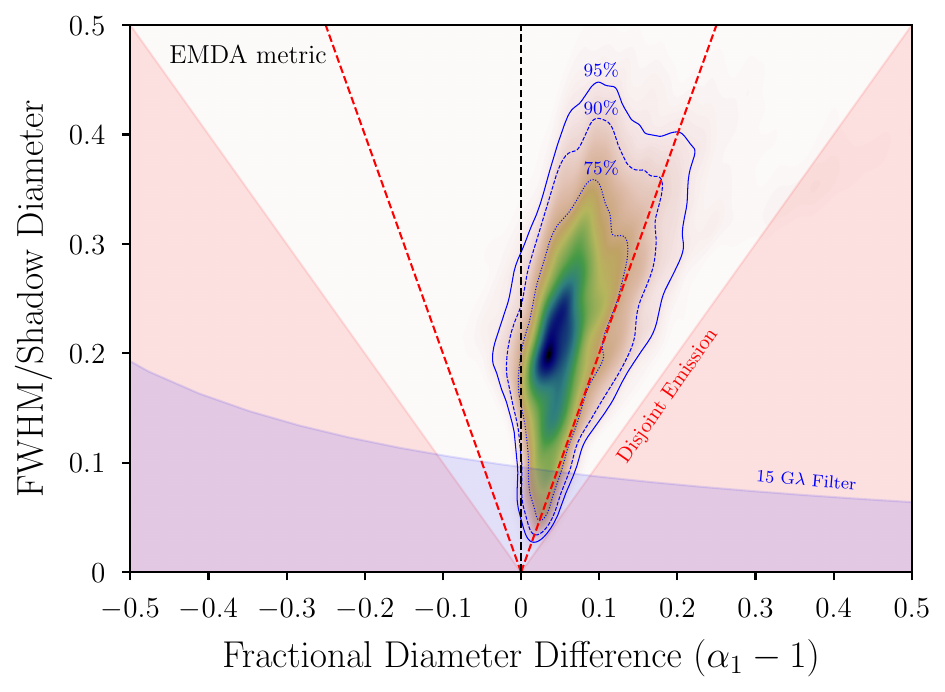}
\end{center}
\vspace*{-3mm}
\caption{Density plots of the photon ring fractional diameter deviation against its fractional width.
The blue dotted, dashed, and solid contour lines correspond, respectively, to $75\%$, $90\%$, and $95\%$ of all models sampled for a given spacetime.
The left panel presents results from the JP image library and the right panel presents the results from the EMDA image library.
Both metric models each total $\sim 10^{5}$ and probe deviations very far away from the Kerr metric, as seen in the broad spread of the contour lines.
Note that the EMDA metric plots exhibit a much narrower spread in diameter difference than the JP metric, owing to these models arising from a particular black hole solution, and in spite of many models having large UPO and ISCO radii.
By contrast, the JP metric contour plot diameter differences are overall broader and in some cases disjoint, owing to the large parametric deformations to the Kerr spacetime introduced by the JP metric's deviation parameters.
}
\label{fig:alpha}
\end{figure*}
\begin{table*}[ht!]
\vspace*{0mm}
\caption{Ranges of the model parameters used to construct the JP metric and EMDA metric image libraries.
Every image has a field of view of $[-20~\rg,\, 20~\rg]$ in both the horizontal and vertical directions and is calculated at a resolution of $128 \times 128$ pixels.
All images are produced at an observing wavelength of $1.3$~mm.
The total number of models in each metric library is $\sim 10^{5}$.
Models with deviation parameters outside the allowed range are omitted (see Appendix.~\ref{Appendix_Char_Radii}).
JP metric models fix two deviation parameters to zero while varying the third parameter.
EMDA metric models vary $b_{\sub *}$ and $\beta_{\sub *}$ simultaneously.
}\vspace*{0mm}
\begin{tabular*}{0.5\textwidth}{c@{\hspace{0.8cm}}cc}
\cmidrule[0.1em]{1-3} 
{\bf Model parameter} & {\bf Library values} & {\bf Parameter description}  \\
\cmidrule[0.05em]{1-3}\morecmidrules\cmidrule[0.05em]{1-3}
$i$ & $15^{\circ},\, 30^{\circ},\, 45^{\circ},\, 60^{\circ},\, 75^{\circ}$ & {observer inclination angle} \\
\cmidrule[0.05em]{1-3}
$a_{\sub *}$ & $-0.9965,\, -0.6509,\, -0.3165,\, 0,\, 0.2940,\, 0.5585,\, 0.7819,\, 0.9428$ & {dimensionless spin parameter} \\
\cmidrule[0.05em]{1-3}
$\epsilon_{\sub 3}$ & $-6,\, -4,\, -2,\, 0,\, 2,\, 4,\, 6,\, 8,\, 10$ & {JP metric deviation parameter} \\
\cmidrule[0.05em]{1-3}
$\alpha_{\sub 13}$ & $-6,\, -4,\, -2,\, 0,\, 2,\, 4,\, 6,\, 8,\, 10$ & {JP metric deviation parameter} \\
\cmidrule[0.05em]{1-3}
$\alpha_{\sub 22}$ & $-2,\, 0,\, 2,\, 4,\, 6,\, 8,\, 10$ & {JP metric deviation parameter} \\
\cmidrule[0.05em]{1-3}
$b_{\sub *}$ & $-0.4,\, -0.2,\, 0,\, 0.2,\, 0.4,\, 0.6,\, 0.8,\, 1$ & {EMDA metric dilaton coupling} \\
\cmidrule[0.05em]{1-3}
$\beta_{\sub *}$ & $0,\, 0.2,\, 0.4,\, 0.6,\, 0.8,\, 1$ & {EMDA metric axion coupling} \\
\cmidrule[0.05em]{1-3}
$\eta$ & $0.05,\, 0.1,\, 0.2$ & {radial velocity profile scale} \\
\cmidrule[0.05em]{1-3}
$n_{r}$ & $0.5,\, 1.0,\, 1.5$ & {radial velocity profile index} \\
\cmidrule[0.05em]{1-3}
$R$ & $5,\, 10$ & {ion-to-electron temperature ratio} \\
\cmidrule[0.05em]{1-3}
$B_{0}~[\mathrm{G}]$ & $5,\, 20,\, 50$ & {magnetic field scale} \\
\cmidrule[0.05em]{1-3}
$\zeta$ & $0.25,\, 0.4$ & {proxy for disk scale height$^1$} \\
\cmidrule[0.05em]{1-3}
\cmidrule[0.1em]{1-3} 
\multicolumn{3}{l}{\small $^{1}$ where in hydrostatic equilibrium and far from $r_{\sub UPO}$ the disk scale height is $h/r=\sqrt{\left( \hat{\gamma} - 1 \right) \zeta}$, i.e., eqn.~\eqref{eq:hr}.}
\label{Tab:JP_library_params}
\end{tabular*}
\vspace*{0mm}
\end{table*}
\subsection{Results from the Image Library \texorpdfstring{$\alpha$}{TEXT}-calibration}\label{Sec:alpha_results}
In Figure~\ref{fig:alpha} we plot the fractional width for each model image against the fractional diameter difference $\alpha_{1} - 1$.
As in the case of the Kerr metric discussed in \citetalias{Ozel2021}, the fractional diameter difference is almost always positive and small, i.e., the bright ring has a slightly larger diameter than the shadow in the vast majority of cases (cf.~Figure 10 of \citetalias{Ozel2021}, but note that the distributions there appear narrower because the panels are separated into positive and negative black hole spins).
This is true for all metric deformation parameters in the various metrics we considered as well as for the details of the plasma model. This is a direct consequence of strong gravitational lensing effects close to $r_{\sub UPO}$ that cause a rapid increase in image brightness near the critical impact parameter and a strong brightness depression interior to it. 

The red filled regions in Figure~\ref{fig:alpha} correspond to images in which the fractional width of the ring is smaller than the fractional diameter difference, while the diagonal dashed lines correspond to half of this value. In other words, models present in that region correspond to those images where the bright ring is displaced from the black hole shadow by more than the ring width. The fact that only a very small fraction of model images lie in that region shows that, even in non-Kerr metrics, ring-like images scale with the diameter of the black-hole shadow and, in the vast majority of cases, are not disjoint from it.

In the case of the non-Kerr images used in calculating the $\alpha$-calibration plots in Figure~\ref{fig:alpha}, we have restricted the analysis to images satisfying two particular requirements.

The first requirement is that a given image must contain at least half of a ring-like feature so that a well-defined diameter is measurable.
As in \citetalias{Ozel2021}, where we defined the fractional coverage of a circle in an image, $\mathcal{F}$, we consider only images with $\mathcal{F}\geq 0.5$.

The second requirement is that the aspect ratio of the non-Kerr black hole shadow, i.e., the ratio of its principal axes, is $0.8 \le \mathcal{R}_{\sub non-Kerr} \le 1.38$. We note that for the Kerr spacetime, this aspect ratio is always constrained to $1 \lesssim \mathcal{R}_{\sub Kerr} \lesssim 1.15$; our requirement corresponds to aspect ratios within $20\%$ of Kerr.
\begin{figure*}[htb!]
\begin{center}
\includegraphics[width=1.0\textwidth]{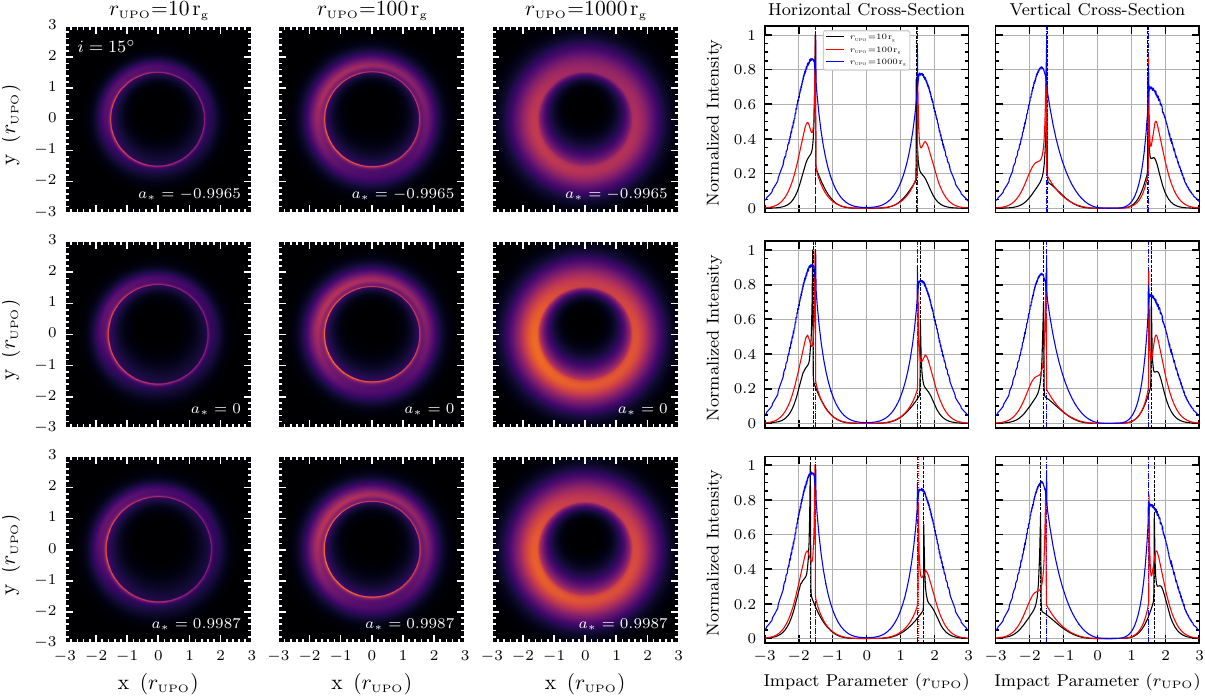}
\end{center}\vspace*{-3mm}
\caption{\footnotesize From left to right, columns 1--3 present image panels of non-Kerr black holes with UPO radii located at $10~\rg$ (column 1), at $100~\rg$ (column 2), and at $1000~\rg$ (column 3).
Each image is individually normalised such that the brightest pixel is of unit intensity.
The color scheme is identical to Figure~\ref{fig:Full_Plasma_Model}.
The observer inclination angle is $15^{\circ}$ in all panels.
The length scale is specified in units of the black hole UPO radius.
All panels in the top, middle, and bottom rows correspond, respectively, to black holes with $a_{\sub *}=-0.9965$ (near-extremal, retrograde spin), $a_{\sub *}=0$ (non-rotating), and $a_{\sub *}=0.9987$ (near-extremal, prograde spin).
Columns 4 and 5 present, respectively, horizontal and vertical intensity cross-sections of their corresponding images in columns 1--3, with horizontal and vertical cross-section pairs normalised such that the largest intensity in the pair is unity.
Vertical black dashed, blue dash-dotted, and red dotted lines delineate the critical impact parameters of non-Kerr black holes with UPO radii of $10~\rg$ (black), $100~\rg$ (red), and $1000~\rg$ (blue).
The peak flux for every image cross-section is always nearly coincident with its associated critical impact parameter, irrespective of the size of the UPO.
}
\label{fig:Vary_UPO_Images_i15}
\end{figure*}

These two requirements are based on the observation that the 1.3~mm image of the M87 black hole is a nearly circular ring~\citep{EHTC2019_PaperVI} and are imposed to eliminate image morphologies that prevent us from using a single radius to characterize images, as is required in the $\alpha$-calibration.
Examples of images eliminated based on the above considerations are presented and discussed in Appendix~\ref{Sec:Appendix_Library}.
\begin{figure*}[htb!]
\begin{center}
\includegraphics[width=1.0\textwidth]{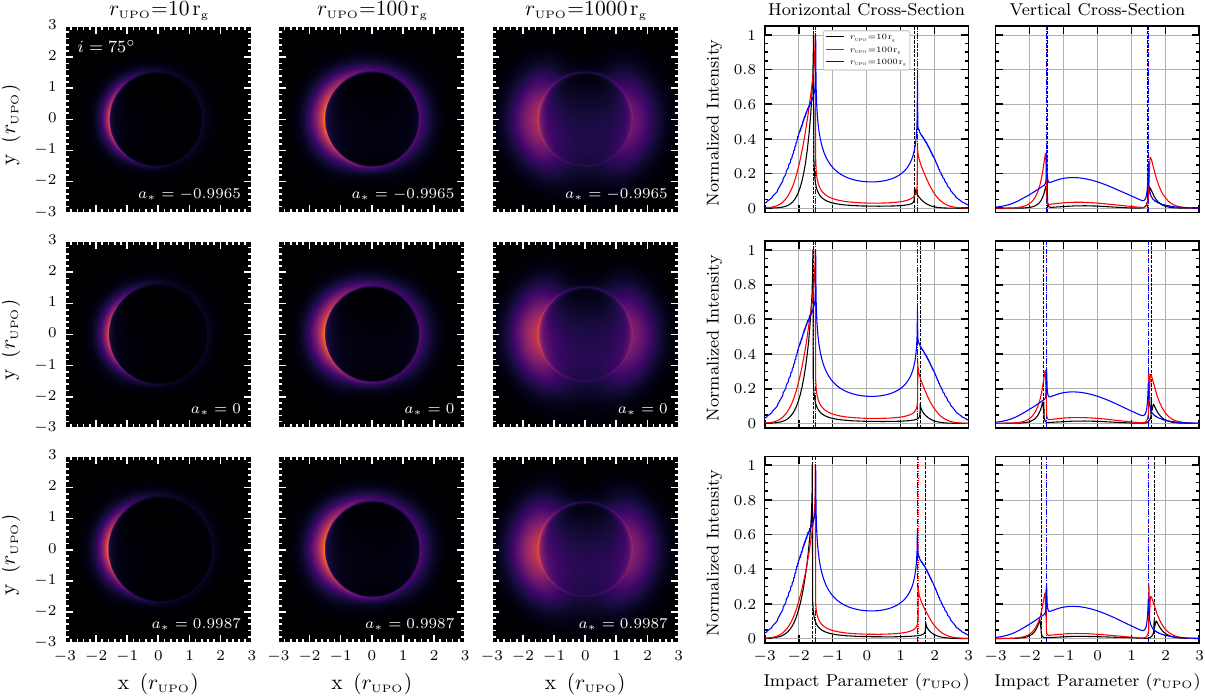}
\end{center}\vspace*{-3mm}
\caption{\footnotesize Same as Figure~\ref{fig:Vary_UPO_Images_i15} but for an observer inclination angle of $75^{\circ}$.
}
\label{fig:Vary_UPO_Images_i75}
\end{figure*}
\subsection{Extreme Deviations from Kerr}
In the choice of values of metric deviation parameters considered in Secs.~\ref{SSec:Library_alpha} and \ref{Sec:alpha_results}, we explored those parameters limited to the ranges shown in Figures~\ref{fig:JP_ISCO}--\ref{fig:EMDA_UPO}.
In particular, for the JP metric, we have restricted ourselves to black holes with UPO radii up to $\sim 5~\rg$.
However, in order to demonstrate the dominant role of gravitational lensing and, in particular, of the critical impact parameter in determining the characteristic features of optically thin black-hole images, Figs.~\ref{fig:Vary_UPO_Images_i15} and \ref{fig:Vary_UPO_Images_i75} explore the images for a number of spacetime parameters with extreme deviations from Kerr. (Note that such deviations are already excluded by the present M87 images).

In particular, these figures show $1.3$~mm images of black holes described by the JP metric with parameters chosen such that the radius of the equatorial circular unstable photon orbit is equal to $10~\rg$, $100~\rg$, and $1000~\rg$.
For these three values of $r_{\sub UPO}$, we choose the dimensionless spin of each of these three black holes to be $-0.9965$, $0$, and $0.9987$, yielding nine images per observer inclination angle and probing a broad range of $a_{*}$ and $r_{\sub UPO}$.
For comparison, for a Kerr black hole $1~\rg < r_{\sub UPO} < 4~\rg$, with $r_{\sub UPO}=3~\rg$ in the case of a Schwarzschild black hole ($a_{\sub *}=0$).
For each pair $(a_{\sub *},r_{\sub UPO})$ the corresponding value of $\alpha_{\sub 13}$ is obtained numerically to an accuracy better than $10^{-16}$.
Table \ref{Tab:Large_UPO_params} presents the numerical values of the parameters that produce these extreme UPO radii.
We note that while these values of $r_{\sub UPO}$ are unphysical in that they are much larger than is constrained by EHT measurements \citep{EHTC2019_PaperVI}, the underlying parameter choices for the metric do not give rise to any pathological behavior or violate any fundamental properties of the spacetime manifold (e.g., closed time-like curves, curvature singularities outside the event horizon, or positive-definite metric determinant values).

The plasma parameters for these extreme examples are: $(\eta,\,n_{r},\,R,\,B_{0},\,\zeta)=(0.5,\,1.5,\,5,\,20~\mathrm{G},\,0.25)$.
Each row in Figures~\ref{fig:Vary_UPO_Images_i15} and \ref{fig:Vary_UPO_Images_i75} presents three images for a fixed value of $a_{\sub *}$, varying the value of $r_{\sub UPO}$, along with their corresponding normalized horizontal and vertical intensity cross-sections.
In order to demonstrate the scaling of the image diameters with the spacetime properties, both the images and the cross-sections are displayed with the impact parameters divided by the corresponding radii of the unstable photon orbits.
Despite such extreme deviations from the Kerr metric which yield enormous UPO radii, one sees that the peak flux in the image is still effectively coincident with the critical impact parameter and a characteristic pronounced central brightness depression is still seen, demonstrating that such image features are effectively independent of the accretion plasma and sensitive to only one property of the spacetime geometry, namely the location of $r_{\sub UPO}$.
\begin{table}[htb!]
\caption{\footnotesize Values of the JP deviation parameter $\alpha_{\sub 13}$ which yield black hole UPO radii of $10~\rg$, $100~\rg$, and $1000~\rg$.
For every UPO radius, $a_{\sub *}$ is chosen as $(-0.9965,\, 0,\, 0.9987)$.
For each calculated value of $\alpha_{\sub 13}$ the resulting ISCO radius is also shown.
By construction, the JP spacetime event horizon radius is the corresponding Kerr value, i.e., $r_{\sub H}/\rg\simeq (1.08359,\,2,\,1.05097)$.}
\centering
\begin{tabular*}{0.5\textwidth}{llcc}
\cmidrule[0.1em]{1-4} 
$r_{\sub UPO} \, \left( \rg \right)$ & $a_{\sub *}$ & $\alpha_{\sub 13}$ & $r_{\sub ISCO} \, \left( \rg \right)$ \\
\cmidrule[0.05em]{1-4}\morecmidrules\cmidrule[0.05em]{1-4}
 & $-0.9965$ & $3.27447\times 10^{2}$ & $3.59824\times 10^{1}$ \\ \cmidrule[0.025em]{3-4}
$10$ & $\phantom{-}0$ & $4.11765\times 10^{2}$ & $3.84384\times 10^{1}$ \\ \cmidrule[0.025em]{3-4}
   & $\phantom{-}0.9987$ & $4.90103\times 10^{2}$ & $4.06180\times 10^{1}$ \\
\cmidrule[0.05em]{1-4}
 & $-0.9965$ & $4.84779\times 10^{5}$ & $1.20952\times 10^{3}$ \\ \cmidrule[0.025em]{3-4}
$100$ & $\phantom{-}0$ & $4.92386\times 10^{5}$ & $1.21871\times 10^{3}$ \\ \cmidrule[0.025em]{3-4}
   & $\phantom{-}0.9987$ & $4.99890\times 10^{5}$ & $1.22773\times 10^{3}$ \\
\cmidrule[0.05em]{1-4}
 & $-0.9965$ & $4.98500\times 10^{8}$ & $3.86751\times 10^{4}$ \\ \cmidrule[0.025em]{3-4}
$1000$ & $\phantom{-}0$ & $4.99249\times 10^{8}$ & $3.87041\times 10^{4}$ \\ \cmidrule[0.025em]{3-4}
   & $\phantom{-}0.9987$ & $4.99998\times 10^{8}$ & $3.87331\times 10^{4}$ \\
\cmidrule[0.1em]{1-4} 
\label{Tab:Large_UPO_params}
\end{tabular*}
\vspace*{3mm}
\end{table}
\begin{figure}[ht!]
\begin{center}
\includegraphics[width=0.47\textwidth]{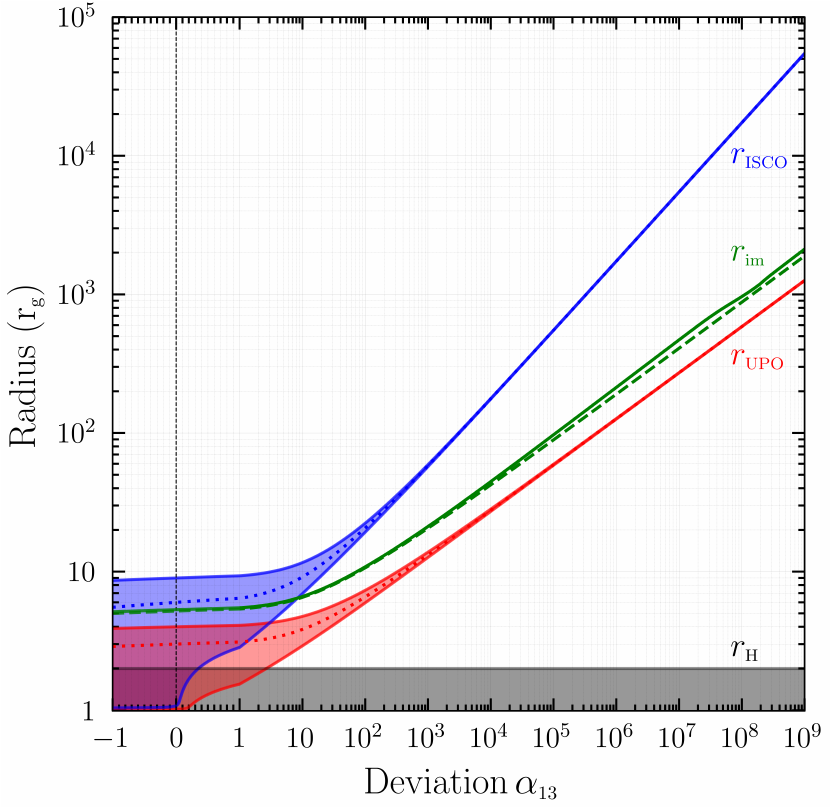}
\end{center}\vspace*{-3mm}
\caption{\footnotesize
Variation of characteristic radii of the JP metric, as a function of the deviation parameter $\alpha_{\sub 13}$.
The three shaded regions show the range of values for the ISCO radii (blue), UPO radii (red), and horizon radii (gray) for the full span of black hole spins explored.
Upper and lower boundaries of the blue and red shaded regions denote extremal negative and positive values of $a_{\sub *}$, respectively.
Their corresponding dotted blue and dotted red curves delineate the case $a_{\sub *}=0$.
The dashed green curve shows the mean radius of the numerically calculated black-hole shadow boundary curve ($d_{\rm sh}/2$).
The solid green curve shows the mean radius of the bright ring feature in the image, as obtained from the image characterization procedure ($r_{\sub im}\equiv d_{\rm im}/2$).
The vertical dotted black line delineates the Kerr metric.
This figure demonstrates that the size of the bright image ring always traces the size of the shadow and that both are determined by the radius of the unstable photon orbit and not by the ISCO or horizon.
The horizontal scale is linear for $\alpha_{\sub 13} < 1$ and logarithmic for $\alpha_{\sub 13} > 1$.
}
\label{fig:Large_UPOs_vary_alpha13}
\end{figure}
\section{Discussion \& Conclusions}\label{Sec:Discussion_and_Conclusions}

In this paper, we explored the dependence of optically thin black-hole images, such as those observed at $1.3$~mm ($230$~GHz) from supermassive black holes with the EHT, on the properties of their spacetimes.
We employed the covariant, semi-analytic model of the accretion plasma we developed in a companion paper~\citepalias{Ozel2021} that is based primarily on conservation laws but allows for a large degree of flexibility in those aspects of the model that depend on the largely unknown mechanism for the turbulent transport of angular momentum.
We applied this covariant, semi-analytic accretion plasma model to both a spacetime which is a solution to the field equations of a non-GR theory of gravity (EMDA), and to a parametrized metric (JP) designed to perturb the Kerr solution and remain agnostic as to the underlying theory of gravity (whilst approximating many other non-GR spacetimes, including EMDA).

We found that, as in the case of the Kerr metric, the dominant characteristics of the images are dictated by the strong gravitational lensing of photons in the vicinity of the unstable photon orbits very close to the black-hole event horizon.
When relativistic Doppler abberational effects due to the azimuthal plasma velocities are not significant, e.g., at lower observer inclinations (where transverse Doppler shifts are negligible), the optically thin images present as narrow rings.
At higher observer inclinations the images can acquire significant brightness asymmetries and become crescent-like.
In all cases the diameters of the images scale with the diameters of the black-hole shadows, with the fractional difference between the two diameters being limited to a small bias.
Moreover, for the vast majority of model images, this fractional difference is not larger than the FWHM of the image itself, i.e., the ring-like images are not disjoint from the boundaries of the shadows. 

Driven by the fact that the inferred shadow size of the black hole in the M87 galaxy is found to be within $\lesssim 20$\% of the value predicted by the Kerr metric~\citep{EHTC2019_PaperVI} and the corresponding constraints on non-Kerr metric parameters are rather stringent~\citep{Psaltis2020}, we have focused so far on metric deviations that are of a similar order.

However, to further underscore the leading role gravitational lensing, coupled with the location of the critical impact parameter and UPO radius, plays in determining the characteristic features of optically-thin black hole images, we also explored black holes with more extreme deviation parameters.
In particular, these black holes, which sampled a broad range of spin parameters, considered UPO radii at $10~\rg$, $100~\rg$, and $1000~\rg$ (see Figs.~\ref{fig:Vary_UPO_Images_i15}--\ref{fig:Vary_UPO_Images_i75}).
Even in the most extreme cases, these figures demonstrate that non-Kerr images remain ring-like or crescent-like with diameters that closely follow the diameters of the black-hole shadows, even though the latter change (with respect to Kerr) by factors of $10$, $100$, and even $1000$.

Figure~\ref{fig:Large_UPOs_vary_alpha13} emphasizes this result by showing various characteristic radii in the JP metric for a very broad range of the deviation parameter $\alpha_{\sub 13}$.
This large range enables us to generate radii for the unstable photon orbit and the ISCO that span $3$--$4$ orders of magnitude in size, while fixing the horizon radius to the Kerr value.
The fact that the image radius closely traces that of the shadow boundary and both track the radius of the unstable photon orbit demonstrates that the latter is the key spacetime characteristic that is accessible to black-hole images.
In contrast, neither the ISCO nor the horizon radii play a significant role in determining the image properties and, therefore, cannot be readily inferred from the observed image sizes.

The examples presented in this study, while varying from marginally different from Kerr to increasingly extreme in deviation and potentially non-physical, serve to provide additional justification for using the diameters of optically-thin back-hole images to infer the sizes of black-hole shadows and, therefore, test the Kerr spacetimes of supermassive black holes.
\section*{Acknowledgements}
Z.,Y.~is supported by a UK Research \& Innovation (UKRI) Stephen Hawking Fellowship and acknowledges partial support from a Leverhulme Trust Early Career Fellowship.
D.\;P.\ and  F.\;\"O.\  acknowledge support from NSF PIRE award OISE-1743747, NSF AST-1715061, and NASA ATP award 80NSSC20K0521.
We thank Monika Mo\'scibrodzka, Lia Medeiros, Mariafelicia de Laurentis, and all members of the Gravitational Physics Working Group of the EHT, for helpful discussions and comments.
We thank the referee for carefully reading the manuscript and providing valuable comments.
This research has made use of NASA's Astrophysics Data System.
\appendix
\section{Characteristic Radii}\label{Appendix_Char_Radii}
\begin{figure*}[htb!]
\begin{center}
\includegraphics[width=0.32\textwidth]{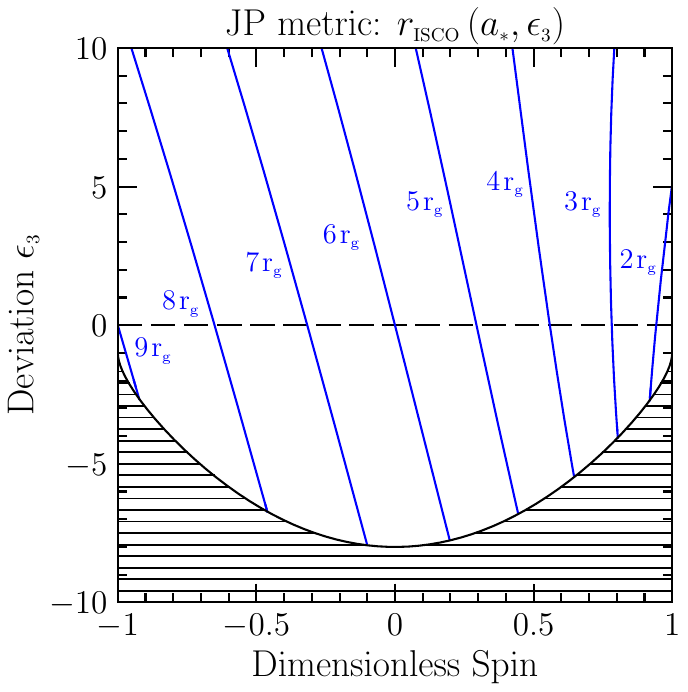}\hfill
\includegraphics[width=0.32\textwidth]{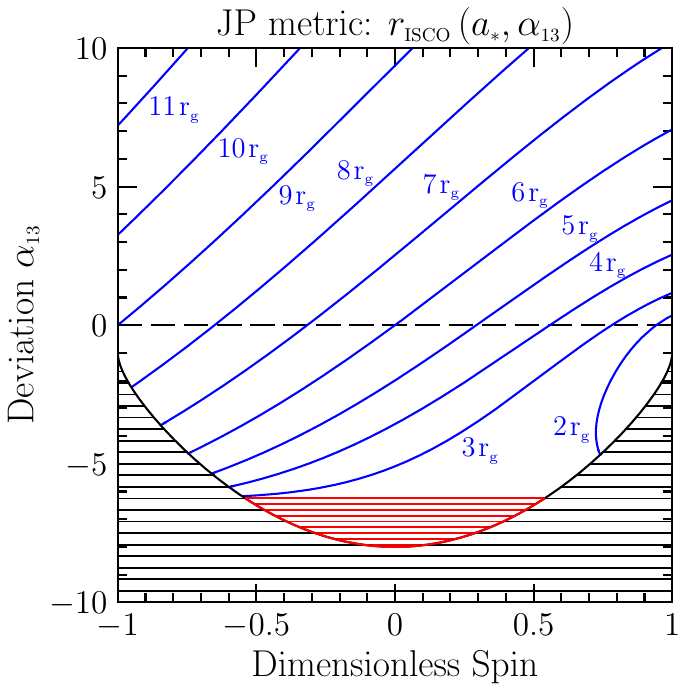}\hfill
\includegraphics[width=0.32\textwidth]{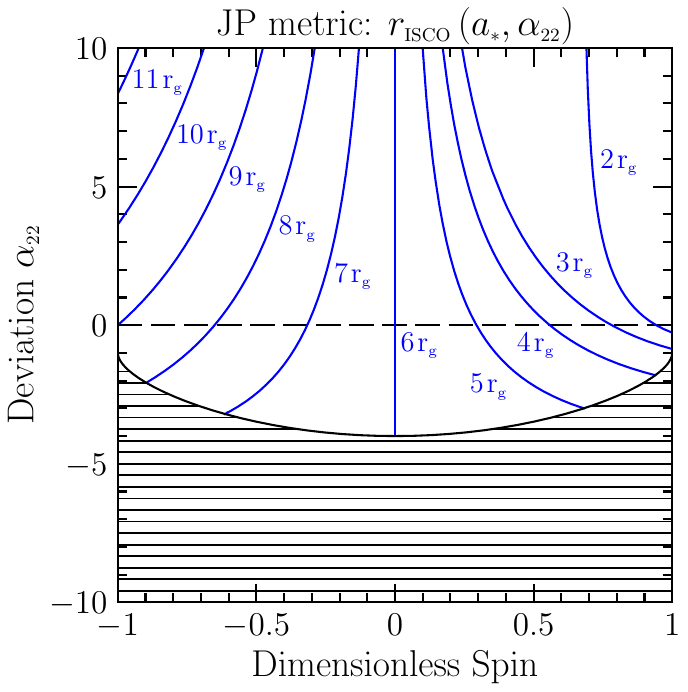}\hfill
\end{center}\vspace*{-3mm}
\caption{\footnotesize Isocontours of the JP metric ISCO radius as a function of dimensionless spin parameter and the deviation parameters $\epsilon_{\sub 3}$ (left), $\alpha_{\sub 13}$ (middle), and $\alpha_{22}$ (right).
In all panels only one deviation parameter is varied, with all others set to zero.
For $\epsilon_{\sub 3}$, the ISCO radius increases as $\epsilon_{\sub 3}$ decreases and $a_{\sub *}$ increases. This holds for all but the highest spins ($a_{\sub *}\gtrsim 0.8$), where the effect of decreasing $\epsilon_{\sub 3}$ yileds an increase in the ISCO radius.
For $\alpha_{\sub 13}$, increasing $a_{\sub *}$ and decreasing the deviation parameter yields a decrease in the ISCO radius.
For $\alpha_{\sub 22}$, the ISCO radius {\it decreases} for $a_{\sub *}<0$ and $\alpha_{\sub 22}$ decreasing, whereas for $a_{\sub *}>0$ the ISCO radius {\it increases} as $\alpha_{\sub 22}$ decreases.
The horizontal dashed line corresponds to a Kerr black hole.
The red shaded region in the middle panel denotes the region where circular equatorial orbits do not exist for radii $\sim2.5\rg$.
The black shaded region delineates the excluded region of the JP parameter space.
These calculations show excellent agreement with Figure~6 of \cite{Johannsen2013b}.}
\label{fig:JP_ISCO}
\end{figure*}
\begin{figure}[htb!]
\begin{center}
\includegraphics[width=0.47\textwidth]{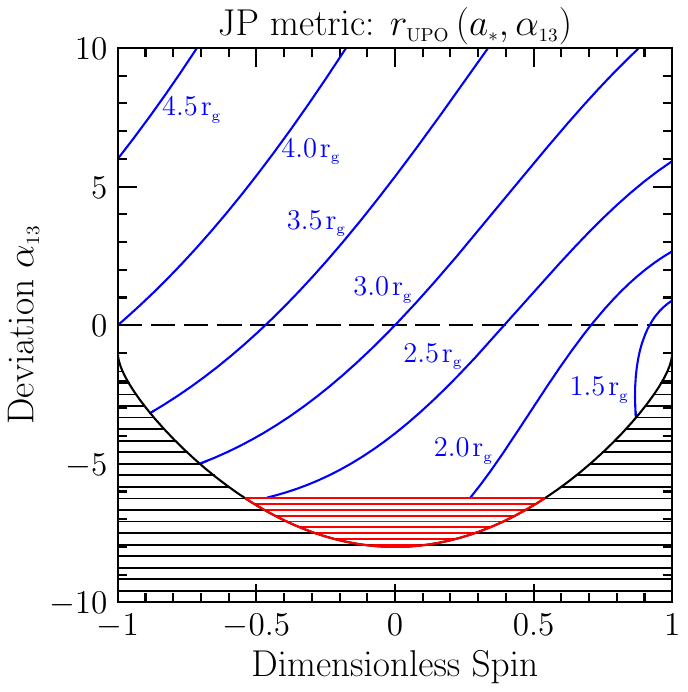}\vspace*{5mm}
\includegraphics[width=0.47\textwidth]{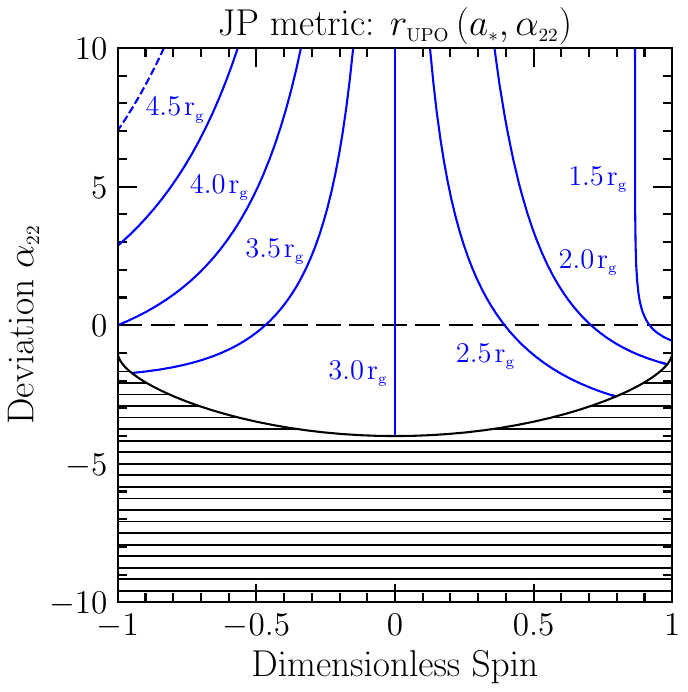}
\end{center}\vspace*{-3mm}
\caption{\footnotesize
Isocontours of circular UPO radius in the JP metric, for the deviation parameters $\alpha_{\sub 13}$ (top) and $\alpha_{\sub 22}$ (bottom).
Parameters and plots as in Figure~\ref{fig:JP_ISCO}.
For $\alpha_{\sub 13}$ the circular UPO radius decreases for decreasing $\alpha_{\sub 13}$ and increasing $a_{\sub *}$.
For $\alpha_{\sub 22}$, when $a_{\sub *}<0$ the circular UPO radius decreases as $\alpha_{\sub 22}$ decreases, whereas for $a_{\sub *}>0$ the circular UPO radius increases as $\alpha_{\sub 22}$ decreases.
Calculations show excellent agreement with Figure~2 of \cite{Johannsen2013c}.
The unlabelled dashed blue isocontour in the upper-left corner of the bottom panel, which was not included in \cite{Johannsen2013c}, denotes $r_{\sub UPO}=5.0~\rg$.}
\label{fig:JP_UPO}
\end{figure}
As discussed in Sec.~3, the covariant plasma model presented in this study depends crucially on the accurate determination of the radii corresponding to the event horizon, UPO, and ISCO.
These calculations depend on specifying the components of the metric tensor of a given spacetime, along with their first and second derivatives (usually in $r$ alone) to machine precision.
We perform our calculations of $r_{\sub H}$, $r_{\sub UPO}$, and $r_{\sub ISCO}$ using quadruple precision arithmetic in \texttt{Fortran} within the \texttt{BHOSS} code \citep{Younsi2012,Younsi2016,Younsi2020}.

The \texttt{BHOSS} code takes the covariant metric tensor of the desired spacetime and numerically calculates all necessary contravariant components and associated metric derivatives to a desired target precision (always $\lesssim 10^{-16}$), which is chosen to be significantly smaller than the geodesic integration tolerance ($\lesssim 10^{-12}$).
These calculations have been validated against a suite of equivalent arbitrary precision routines in \texttt{Mathematica}, confirming their accuracy across the entire range of metric parameters considered in this study.
In the following subsections the equations defining the characteristic radii are described.
Results of their solution for different spacetimes, which are in excellent agreement with previously-published results, are also presented.
\subsection{Calculation of \texorpdfstring{$r_{\sub H}$}{TEXT}}
As described in \cite{Johannsen2013b}, a sufficiently general definition of the event horizon of a general black hole may be encompassed by the condition $r=H( \theta )$, which is defined by the condition:
\begin{equation}
g^{rr}( r,\theta) - 2 g^{r\theta} ( r,\theta ) \left[ \frac{{\rm d}H(\theta )}{{\rm d}\theta} \right] + g^{\theta\theta}( r,\theta ) \left[ \frac{{\rm d}H(\theta )}{{\rm d}\theta} \right]^{2} = 0 \,, 
\label{eq:rH_general}
\end{equation}
where $g^{r\theta}=0$ for all metrics investigated in this study.
Equation \eqref{eq:rH_general} is solved numerically through discretisation of the zenith derivative using a backward finite-difference approximation, starting on the polar axis ($\theta=0$) and iterating until $\theta=\pi/2$.

For the Kerr black hole the (outer) event horizon radius is given by the larger solution to $\Delta=0$, namely:
\begin{equation}
r_{\sub H,Kerr} = \rg \left( 1 + \sqrt{1 - a_{\sub *}^{2}} \right) \,.
\end{equation}
The EMDA black hole event horizon radius is given by:
\begin{equation}
r_{\sub H,EMDA} = \rg \left[ \left(1+b_{\sub *}\right) + \sqrt{ \left(1 + \beta_{b}^{2}\right) \left(1 + b_{\sub *}\right)^{2}  - a_{\sub *}^{2} } \right] \,.
\end{equation}
Given a choice of metric parameters which yield a ``well-behaved'' metric, when $\alpha_{\sub 52}=0$ the event horizon of the JP metric is precisely the Kerr event horizon.
In this study the event horizon radii for the EMDA and JP metrics are both calculated fully numerically via eq.~\eqref{eq:rH_general}, with the analytic expressions above serving as a check of the accuracy of the numerical root finding algorithms used to determine $r_{\sub H}$.
\subsection{Calculation of \texorpdfstring{$r_{\sub UPO}$}{TEXT}}
Consider a general static and axisymmetric metric:
\begin{equation}
\mathrm{d}s^{2}=g_{tt}\mathrm{d}t^{2}+2g_{t\phi}\mathrm{d}t \, \mathrm{d}\phi+g_{rr}\mathrm{d}r^{2}+g_{\theta\theta}\mathrm{d}\theta^{2}+g_{\phi\phi}\mathrm{d}\phi^{2}\,.
\end{equation}
The Lagrangian equations of motion $2\mathcal{L}=g_{\mu\nu}\dot{x}^{\mu}\dot{x}^{\nu}$ for this metric yield the constants $p_{t}\equiv-E=g_{tt}\dot{t}+g_{t\phi}\dot{\phi}$ and $p_{\phi}\equiv L_{\rm z}=g_{t\phi}\dot{t}+g_{\phi\phi}\dot{\phi}$, where an overdot denotes differentiation with respect to proper time ($\tau$).
For a fluid particle, the normalization condition $u_{\mu}u^{\mu}=-1$, together with the assumption of circular orbits in the equatorial plane, i.e., $\theta=\pi/2$ and $\dot{\theta}=0$, when paired with the radial equation of geodesic motion yields:
\begin{subequations}
\begin{eqnarray}
\rho^{2}\dot{t} &=& \left( g_{t\phi} L_{\rm z} + g_{\phi\phi} E \right) \,, \label{eqn:tdot} \\
\rho^{2}\dot{\phi} &=& -\left( g_{tt} L_{\rm z} + g_{t\phi} E \right) \,, \label{eqn:phidot} \\
g_{rr}\dot{r}^{2} &=& -\left( g_{tt} + 2 g_{t\phi}\Omega + g_{\phi\phi}\Omega^{2} \right) \dot{t}^{2} - 1 \,, \label{eqn:rdot}
\end{eqnarray}
\end{subequations}
where $\Omega:=\dot{\phi}/\dot{t}$ and $\rho^{2}\equiv (g_{t\phi})^{2}-g_{tt}\,g_{\phi\phi}$.

Circular equatorial orbits imply $\dot{r}=\ddot{r}=0$, from which one obtains the condition $g_{tt,r}+2g_{t\phi,r}\Omega+g_{\phi\phi,r}\Omega^{2}=0$, where ${f}_{,\mu}\equiv \partial_{\mu}f:=\partial f/\partial x^{\mu}$.
Solving this condition yields the orbital angular velocity of circular orbits in terms of first derivatives of the metric:
\begin{equation}
\Omega = \frac{-g_{t\phi,r} \pm \sqrt{(g_{t\phi,r})^{2}-g_{tt,r} \, g_{\phi\phi,r}}}{g_{\phi\phi,r}} \,,
\label{eqn:Omega_circ}
\end{equation}
where the positive sign denotes orbits which are co-rotating with a prograde spinning black hole ($a_{\sub *}>0$).
Counter-rotating orbits (negative sign in eq.~\eqref{eqn:Omega_circ}) are not considered separately in this study, as we allow for the black-hole spin to be negative.
The condition $\dot{r}=0$ yields:
\begin{equation}
u^{t} = \left( -g_{tt} - 2 g_{t\phi}\Omega - g_{\phi\phi}\Omega^{2} \right)^{-1/2} \,,
\end{equation}
from which the energy and angular momentum of a particle in circular orbit in the equatorial plane may be written as:
\begin{subequations}
\begin{eqnarray}
E &=& -u^{t}\left( g_{tt} + g_{t\phi} \Omega \right) \,, \label{eqn:E_circ} \\
L_{z} &=& u^{t}\left( g_{t\phi} + g_{\phi\phi} \Omega \right) \,. \label{eqn:Lz_circ}
\end{eqnarray}
\end{subequations}
The UPO radius is given by the smallest real value of $r\ge r_{\sub H}$ for which, in the limit $r \rightarrow r_{\sub UPO}$, both $E^{-1}$ and $L_{\rm z}^{-1}$ tend to zero. Consequently, $r_{\sub UPO}$ is obtained from $(u^{t})^{-1}=0$ and is calculated numerically as the solution of the equation:
\begin{equation}
\left(g_{tt} + 2 g_{t\phi}\Omega + g_{\phi\phi}\Omega^{2}\right)|_{r=r_{\sub UPO}} = 0 \,.
\end{equation}
\subsection{Calculation of \texorpdfstring{$r_{\sub ISCO}$}{TEXT}}
Substituting eqs.~\eqref{eqn:tdot} and \eqref{eqn:phidot} into eq.~\eqref{eqn:rdot} yields, upon simplification:
\begin{equation}
\rho^{2}g_{rr}\dot{r}^{2} = g_{\phi\phi}E^{2} + 2g_{t\phi}E L_{\rm z} + g_{tt} L_{\rm z}^{2} - \rho^{2} \,.
\label{eqn:ISCO_part1}
\end{equation}
Recalling the condition $\dot{r}=\ddot{r}=0$, upon application to \eqref{eqn:ISCO_part1}, one obtains:
\begin{equation}
g_{\phi\phi,rr}E^{2} + 2g_{t\phi,rr}E L_{\rm z} + g_{tt,rr} L_{\rm z}^{2} - (\rho^{2})_{,rr}=0 \,.
\label{eqn:r_ISCO}
\end{equation}
The ISCO radius is then obtained numerically as the smallest real solution of \eqref{eqn:r_ISCO} satisfying $r_{\sub ISCO}\ge r_{\sub UPO}$.

\subsection{Characteristic radii for JP and EMDA spacetimes}
This subsection presents figures of isocontours of UPO and ISCO radii for the JP and EMDA spacetimes, as a function of different metric parameters and $a_{\sub *}$.
The excluded regions for the JP spacetime are given by: $\epsilon_{\sub 3} \le -r_{\sub +}^{3}$, $\alpha_{\sub 13}\le -r_{\sub +}^{3}$, and $\alpha_{\sub 22} \le -r_{\sub +}^{2}$, where $r_{\sub +}\equiv 1+\sqrt{1-a_{\sub *}^{2}}$.
For the EMDA spacetime an exclusion region exists when $\beta_{\sub *}=0$, and is given by: $b_{\sub *} \le -(1-a_{\sub *}^{2})/2$.

\begin{figure*}[htb!]
\begin{center}
\includegraphics[width=0.32\textwidth]{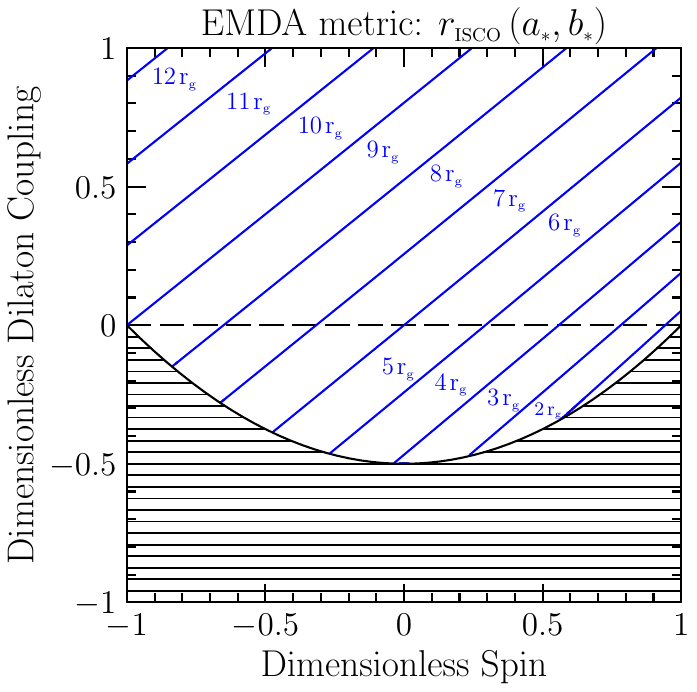}\hfill
\includegraphics[width=0.32\textwidth]{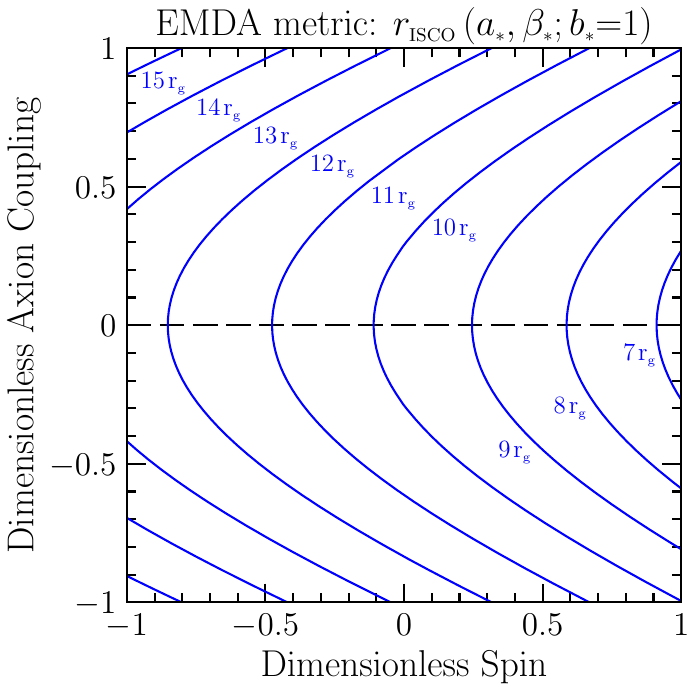}\hfill
\includegraphics[width=0.32\textwidth]{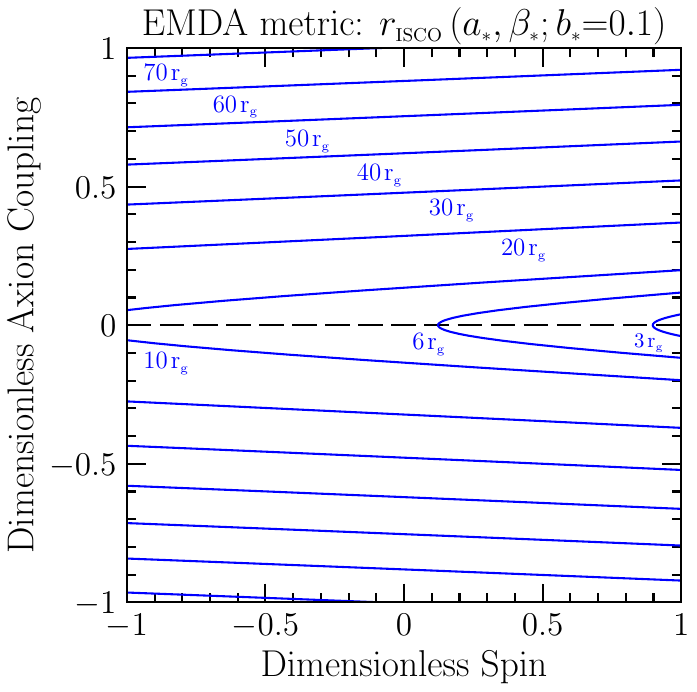}\hfill
\end{center}\vspace*{-3mm}
\caption{\footnotesize
Isocontours of ISCO radius in the EMDA metric, for the dimensionless dilaton coupling (left), and for the dimensionless axion coupling with fixed dilaton coupling values of $1$ (middle) and $0.1$ (right).
The horizontal dashed line corresponds to a Kerr black hole.
The black shaded region delineates the excluded region of the EMDA parameter space when $\beta_{\sub *}=0$.
In the left panel, the dilaton coupling parameter is varied whilst the axion coupling is set to zero.
The ISCO radius decreases for decreasing dilaton coupling and increasing $a_{\sub *}$.
For the middle and right panels, the EMDA ISCO radius dependence on the axion coupling is symmetric under $\beta_{\sub *} \rightarrow -\beta_{\sub *}$, exhibiting the same trend as the dilaton case.
Letting $b_{\sub *}\rightarrow 0$ causes $\beta_{b}\rightarrow \infty$ to diverge, resulting in all characteristic radii of the (non-zero axion coupling) EMDA solution to increase rapidly.}
\label{fig:EMDA_ISCO}
\end{figure*}

\begin{figure*}[htb!]
\begin{center}
\includegraphics[width=0.32\textwidth]{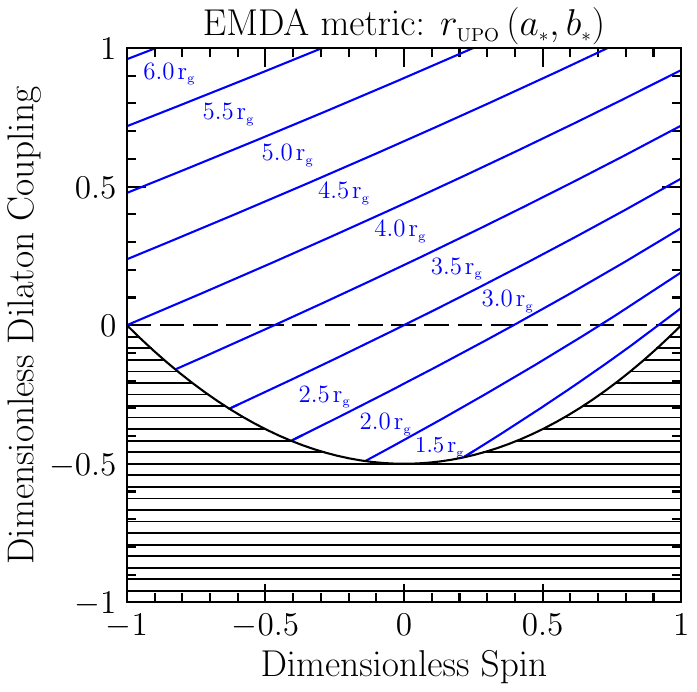}\hfill
\includegraphics[width=0.32\textwidth]{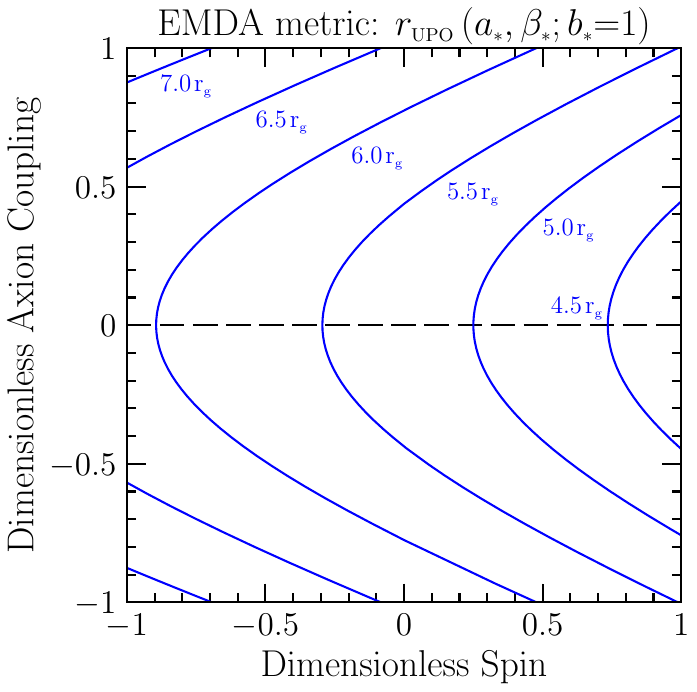}\hfill
\includegraphics[width=0.32\textwidth]{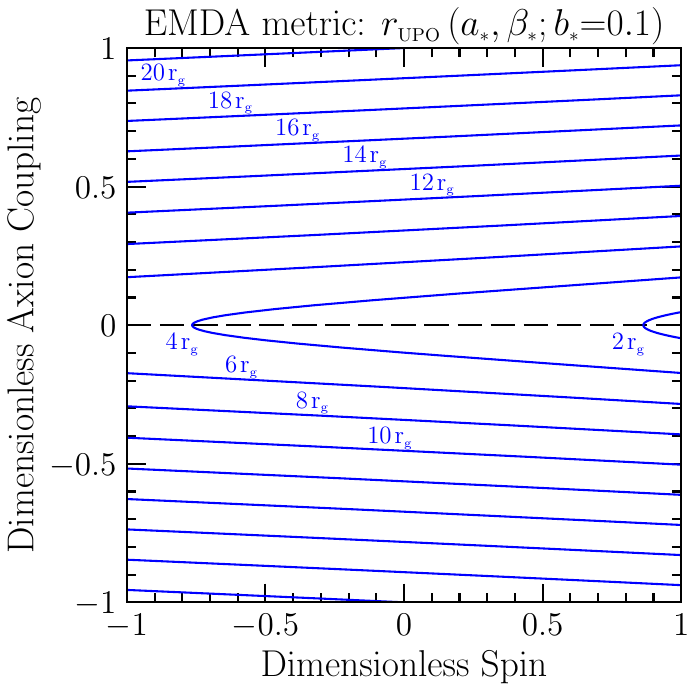}\hfill
\end{center}\vspace*{-3mm}
\caption{\footnotesize
Isocontours of circular UPO radius in the EMDA metric, for the dimensionless dilaton coupling (left), and for the dimensionless axion coupling with fixed dilaton coupling values of $1$ (middle) and $0.1$ (right).
Plots and trends are as described in Figure~\ref{fig:EMDA_ISCO}.}
\label{fig:EMDA_UPO}
\end{figure*}

In the case of the JP metric, the parameter $\alpha_{\sub 52}$ affects only the $g_{rr}$ metric component, which does not affect the UPO and ISCO radii, thus fixing $\alpha_{\sub 52}=0$ ensures the JP and Kerr horizon radii are always coincident.
The $\epsilon_{3}$ parameter only affects, albeit weakly, the ISCO radius (see Fig.~\ref{fig:JP_ISCO}).
JP metric isocontours of ISCO and UPO radius, for varying values of one deviation parameter and spin (with all other parameters set to zero) are shown in Figs.~\ref{fig:JP_ISCO} and \ref{fig:JP_UPO}.
These results are in excellent agreement with \cite{Johannsen2013c}, where we note the shaded red region in the $\alpha_{\sub 13}$ UPO plot of Figure~\ref{fig:JP_UPO}, which is absent in Figure~2 of \cite{Johannsen2013c} since they considered a truncated Taylor expansion approximation of the underlying equations rather than the full numerical solution employed here.
This was subsequently rectified in Figure~6 of \cite{Johannsen2013b}.
For the EMDA metric, the cases $\beta_{\sub *}=0$, $b_{\sub *}=1$, and $b_{\sub *}=0.1$ are shown in Figs.~\ref{fig:EMDA_ISCO} and \ref{fig:EMDA_UPO}.

\section{Free-fall plasma 4-velocities}\label{Appendix:free_fall}
This section presents the derivation of 4-velocity profiles describing free-falling particle geodesics in the EMDA and JP spacetimes.
Test particle motion in arbitrary spacetime geometries may be described in a straightforward manner as follows.
Static, axisymmetric spacetimes which are integrable admit two Killing vectors corresponding to the particle energy relative to infinity, $E\equiv-p_{t}$, and the component of the particle angular momentum about the axis of symmetry of the black hole, $L_{\rm z}\equiv p_{\phi}$.

In addition to these symmetries, which arise from the stationary and axisymmetric nature of these spacetimes, like the Kerr metric, such spacetimes admit a rank-2 Killing tensor which acts as a separation constant in the Hamilton-Jacobi equation, i.e., it yields a Carter constant, $Q$ \citep{Carter1968,Walker1970,Benenti1979,Papadopoulos2018}.
Such spacetimes therefore admit three constants of motion and by further utilising the conservation of rest mass, $\mu$, of the particle itself, i.e., $\mu^{2} \equiv -g^{\alpha\beta}p_{\alpha}p_{\beta}$ (where $p_{\beta}$ is the particle's canonical 4-momentum), the system of equations describing the particle's motion is well-determined and expressible as a system of four first-order ordinary differential equations (ODEs).
This reduces the geodesic motion to a quadrature problem.

\subsection{Free-fall in the EMDA spacetime}
It is a straightforward albeit lengthy exercise to determine the EMDA metric equations of motion from the Hamilton-Jacobi equation.
After specifying an appropriate ansatz for the separation constant of the Hamilton-Jacobi equation, simplification of the resulting equations of motion of a test particle in the EMDA spacetime yields:
\begin{subequations}
\begin{eqnarray}
\mu \widehat{\Sigma} \left(\frac{{\rm d} t}{{\rm d} \tau}\right) &=& -a W \left( a E W\sin^{2}\theta - L_{\rm z}\right) + \frac{\delta \, \mathcal{P}}{\widehat{\Delta}} \,, \hspace*{8mm} \\
\mu \frac{\widehat{\Sigma}}{c} \left(\frac{{\rm d} r}{{\rm d} \tau}\right) &=& \pm \sqrt{\widehat{\mathcal{R}}(r)} \,, \\
\mu \frac{\widehat{\Sigma}}{c} \left(\frac{{\rm d} \theta}{{\rm d} \tau}\right) &=& \pm \sqrt{\widehat{\Theta} \left(\theta \right)} \,, \label{eq:zenith_EMDA_particle}\\
\mu \frac{\widehat{\Sigma}}{c} \left(\frac{{\rm d} \phi}{{\rm d} \tau}\right) &=& -\left( a E W - L_{\rm z}^{2} \csc^{2}\theta \right) + \frac{a \, \mathcal{P}}{\widehat{\Delta}} \,,
\end{eqnarray}
\label{eq:EMDA_particle}
\end{subequations}
and where
\begin{subequations}
\begin{eqnarray}
\widehat{\mathcal{R}}(r) :=&& \mathcal{P}^{2}-\widehat{\Delta}\left[ \mu^{2}\left(r^{2}-2br - \beta^{2}\right) + \mathcal{Q} \right] \,, \\
\widehat{\Theta} \left( \theta \right) :=&& - \mu^{2}\left[ a^{2}\cos^{2}\theta + \rg^{2}\beta_{b}\left(\beta_{b}-2a_{\sub *}\cos\theta\right) \right] \nonumber \\
&&-\frac{\left(L_{\rm z} - a E W \sin^{2}\theta\right)^{2}}{\sin^{2}\theta} + \mathcal{Q} \,, \\
\mathcal{P} :=&& E\delta - a L_{\rm z} \,, \\
\mathcal{Q} :=&& Q + \left(L_{\rm z}-a E\right)^{2} \,.
\end{eqnarray}
\end{subequations}
The velocities of radially free-falling test particles are obtained when the constants of motion are $E=\mu$, $L_{\rm z}=0$, and $Q=0$. 
Upon substitution into eqs.~\eqref{eq:EMDA_particle}, one obtains:
\begin{subequations}
\begin{eqnarray}
\widehat{\Sigma} \left(\frac{{\rm d} t}{{\rm d} \tau}\right)_{\rm ff} &=& -a^{2}W^{2}\sin^{2}\theta + \frac{\delta^{2}}{\widehat{\Delta}} \,, \\
\frac{\widehat{\Sigma}}{c} \left(\frac{{\rm d} r}{{\rm d} \tau}\right)_{\rm ff} &=& -\sqrt{\delta^{2} - \widehat{\Delta}\left( \delta - \beta^{2} \right)} \,, \\
\frac{\widehat{\Sigma}}{c} \left(\frac{{\rm d} \theta}{{\rm d} \tau}\right)_{\rm ff} &=& \pm\sqrt{-\beta\left[\beta + \frac{X}{b^{2}} + \frac{\beta X^{2}}{a^{2}b^{4}\sin^{2}\theta} \right]} \,, \hspace*{5mm} \label{eq:zenith_EMDA_freefall}\\
\frac{\widehat{\Sigma}}{c} \left(\frac{{\rm d} \phi}{{\rm d} \tau}\right)_{\rm ff} &=& a\left( -W + \frac{\delta}{\widehat{\Delta}} \right) \,,
\end{eqnarray}\label{eq:EMDA_freefall}
\end{subequations}
where $X\equiv 2 \, a \, b \, \rg \cos\theta +\beta(b^{2}-\rg^{2})$ and the subscript ``ff'' denotes free fall.
In the above, the negative root of the radial motion is taken, corresponding to free fall onto the black hole.
Note, however, that even in the free fall case the zenith motion does not vanish for the full EMDA metric, due to the axion field coupling parameter, i.e., the aforementioned ansatz fails since the axion field coupling breaks the integrability of the motion.
In this study we therefore set $\beta_{\sub *}=0$ when considering free fall motion in the EMDA spacetime.
This ensures eqs.~\eqref{eq:zenith_EMDA_particle} \& \eqref{eq:zenith_EMDA_freefall} vanish when $L_{\rm z}=Q=0$, thereby constraining the motion of a test particle to be confined to a two-dimensional plane, as required.
The standard Kerr expressions are recovered in the limit $b\rightarrow 0$.

\subsection{Free fall in the JP spacetime}
Similar to the previous subsection for the EMDA spacetime, the equations of motion for particles in the JP metric (see Johannsen, 2015) may be re-expressed in the free-falling regime as:
\begin{subequations}
\begin{eqnarray}
\widetilde{\Sigma} \left(\frac{{\rm d} t}{{\rm d} \tau}\right)_{\rm ff} &=& -a^{2} \sin^{2} \theta +\frac{\left( r^{2} + a^{2} \right)^{2} A_{\sub 1}^{2}}{\Delta} \,, \\
\frac{\widetilde{\Sigma}}{c} \left(\frac{{\rm d} r}{{\rm d} \tau}\right)_{\rm ff} &=& -\sqrt{A_{\sub 5}} \ \sqrt{\mathcal{K} + 2\,\rg r \left( r^{2} + a^{2} \right)} \,,\hspace*{3mm} \\
\frac{\widetilde{\Sigma}}{c} \left(\frac{{\rm d} \theta}{{\rm d} \tau}\right)_{\rm ff} &=& 0 \,, \\
\frac{\widetilde{\Sigma}}{c} \left(\frac{{\rm d} \phi}{{\rm d} \tau}\right)_{\rm ff} &=& a\left[ -1 + \frac{\left(r^{2} + a^{2} \right) A_{\sub 1} A_{\sub 2}}{\Delta} \right] \,,
\end{eqnarray}
\end{subequations}
where $\mathcal{K} \equiv \left(A_{\sub 1}^{2} - 1 \right)\left(r^{2} + a^{2}\right)^{2}-\Delta \left( \widetilde{\Sigma} - \Sigma \right)$.
Note that for the JP metric the existence of three constants of motion is guaranteed by construction.
The standard Kerr metric expressions are recovered when all deviation parameters are zero.
\section{Images excluded from the \texorpdfstring{$\alpha$}{TEXT}-calibration}\label{Sec:Appendix_Library}
\begin{figure*}[htb!]
\begin{center}
\includegraphics[width=0.98\textwidth]{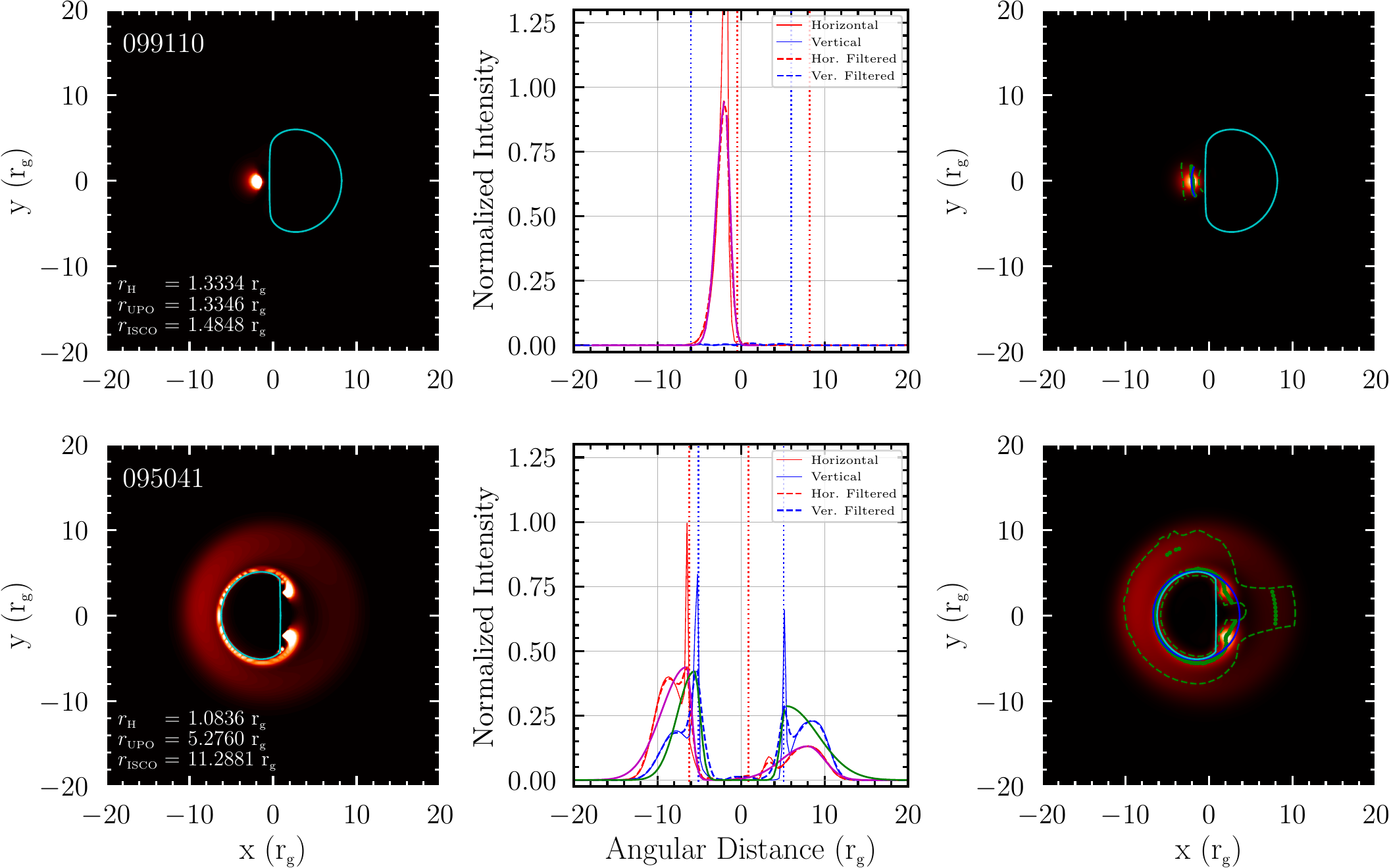}
\end{center}
\vspace*{-3mm}
\caption{\footnotesize Examples of images excluded from the $\alpha$-calibration analysis summarised in Figure~\ref{fig:alpha}, which the $\alpha$-calibration would artificially characterize as disjoint.
The top row presents an example of a model where $\mathcal{F} \simeq 0.13$, the smallest fraction of all models in the image library.
The bottom row presents a model with $\mathcal{F}=1$ and $\mathcal{R}_{\sub non-Kerr}$ outside the Kerr aspect ratio constraints.
The left column presents these example images, for an $a_{\sub *}=0.9428,\,i=75^{\circ},\,\alpha_{\sub 22}=10$ black hole (top) and an $a_{\sub *}=-0.9965,\,i=15^{\circ},\,\alpha_{\sub 22}=10$ black hole (bottom).
The middle column presents horizontal and vertical intensity cross-sections of these images, together with these cross-sections, as obtained from the filtered images (right column).
The black hole shadow boundary curve is delineated by the cyan curve in the left and right columns.
The blue line in the rightmost panels denotes the contour of median radii where the flux is non-negligible and FWHM radii are denoted by dashed green lines.
The large quadrupole moments of these black hole produce highly non-circular ``D-shaped'' shadows that cannot be characterized by a single diameter.
}
\label{fig:JP_library_filter_examples}
\end{figure*}
As mentioned in Sec.~\ref{Sec:alpha_results}, model images that deviate significantly from circular shapes are excluded from the $\alpha$-calibration analysis presented therein. Specifically, such images do not present at least half circles, as characterized by $\mathcal{F}<0.5$, or fall outside of the constraint $0.8 \le \mathcal{R}_{\sub non-Kerr} \le 1.38$. In these cases, even though the ring-like shapes still closely follow the shadow boundaries, a characterization based on a circular shape fails to correctly capture the true relationship between the two. In other words, violating the two conditions listed above prevent us from using a single radius to characterize these images, as is required in the $\alpha$-calibration approach, which is based on the observation that the image in the M87 black hole is nearly circular. 

The models with those characteristics typically have large dimensionless spin magnitudes, are viewed close to edge-on $(i=75^{\circ})$, and arise from the largest values of deviation parameters.
Many of these models have large values of $\alpha_{\sub 22}$, as discussed in Figure~\ref{fig:JP_library_filter_examples}, rendering the shadow boundary curve highly prolate and the images asymmetric. As can be seen in the rightmost panels of this figure, the present characterisation algorithm is not suited to dealing with highly non-circular images, which are not consistent with observations. If future observations of other black holes reveal highly non-circular images, the calibration can easily be extended to incorporate such shapes. 
\section{Non-GR Radiative Transfer Code Comparison}\label{Sec:GRRT_Code_Comparison}
Cross-code verification of time-dependent GRMHD and of radiative transfer algorithms in the Kerr metric have been reported in earlier publications by the EHT collaboration~\citep{Porth2019,Gold2020}. In this Appendix, we provide a cross-code verification of the numerical implementation for radiative transfer in non-GR metrics and for the plasma model that we employ in this study.

For the purposes of this verification, we employ two radiative transfer algorithms, which were designed specifically to handle general spacetimes that may not possess the symmetries of the Kerr metric. 

The first algorithm is described in \cite{Psaltis2012}.
It employs the Killing vectors related to the stationarity and axisymmetry of a general spacetime but integrates the second-order geodesic equations for the remaining spacetime coordinates using a fourth order Runge-Kutta-Fehlberg integrator with adaptive step size control.

The second algorithm is described in \cite{Younsi2012,Younsi2016}.
It integrates the second-order geodesic equations for all spacetime coordinates and does not make use of any symmetries of the spacetime.
It solves these equations using fourth-order, sixth-order, and eighth-order Runge-Kutta-Fehlberg integration routines with adaptive step size control, as well.

\begin{figure*}[htb!]
\begin{center}
\includegraphics[width=0.34\textwidth]{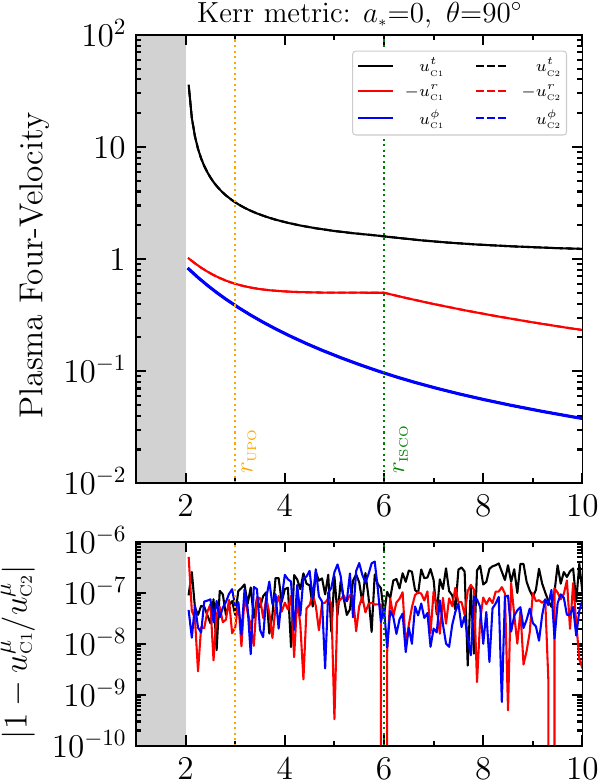}\hfill
\includegraphics[width=0.31\textwidth]{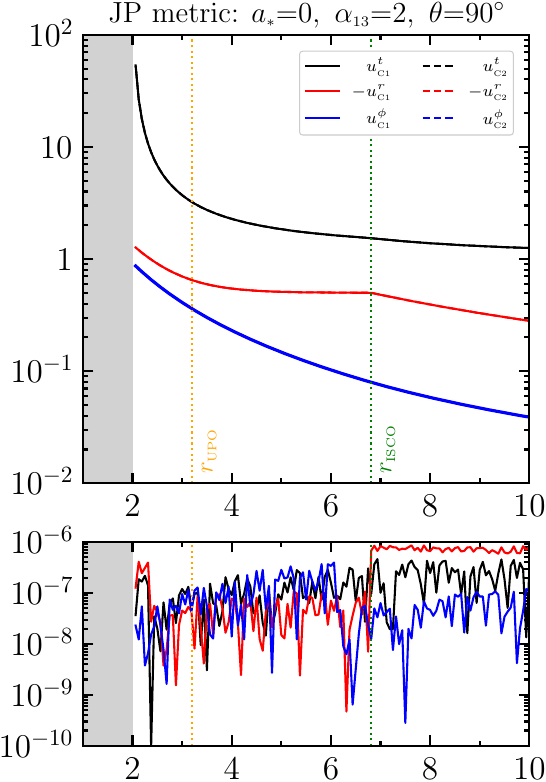}\hfill
\includegraphics[width=0.31\textwidth]{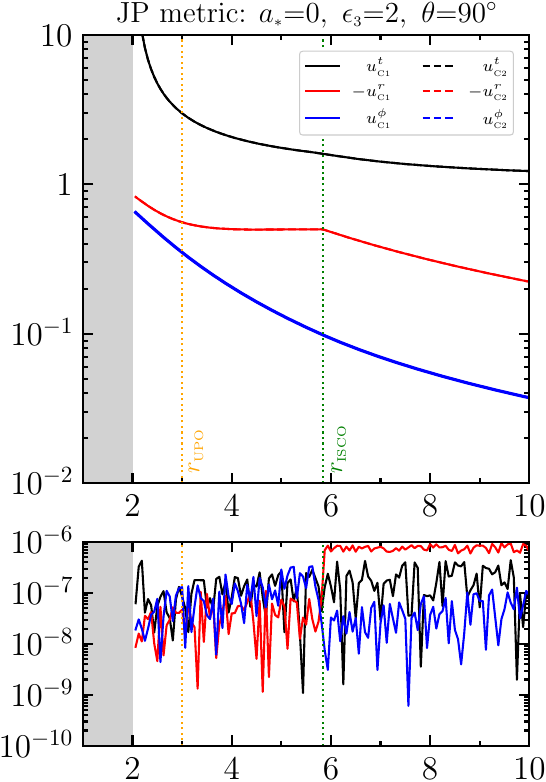}\hfill
\end{center}
\vspace*{-3mm}
\begin{center}
\includegraphics[width=0.34\textwidth]{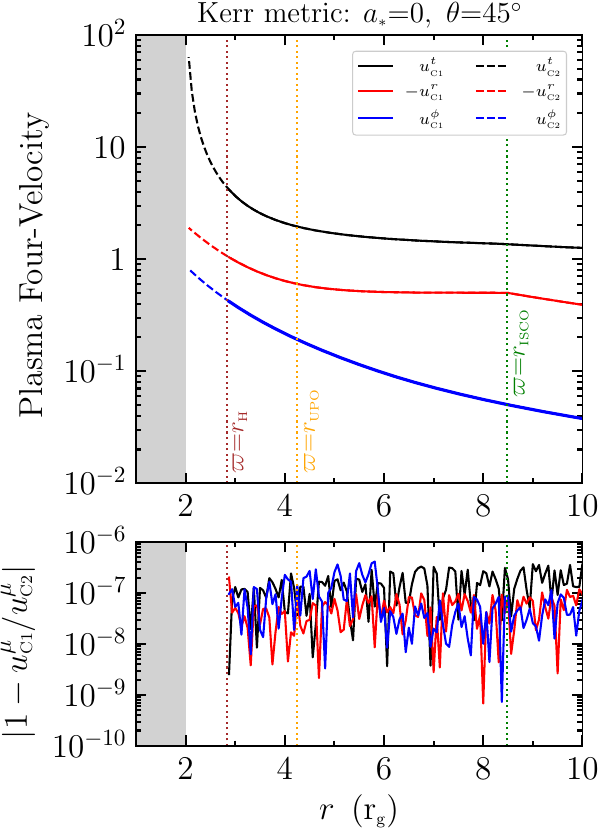}\hfill
\includegraphics[width=0.31\textwidth]{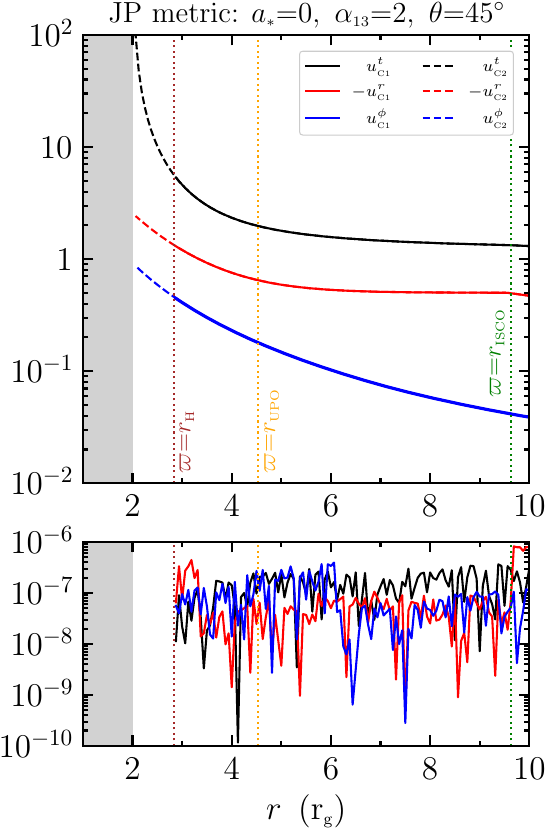}\hfill
\includegraphics[width=0.31\textwidth]{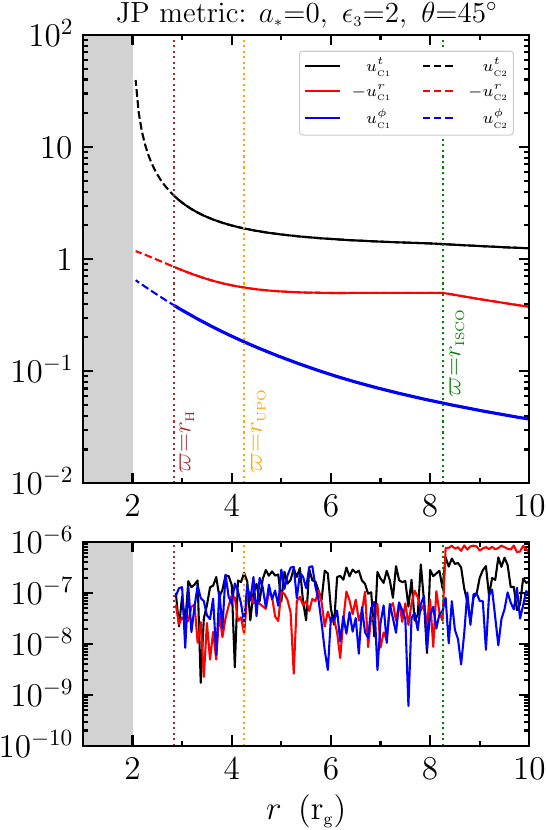}\hfill
\end{center}\vspace*{-3mm}
\caption{\footnotesize The three non-zero components of the plasma velocities in the model described in Sec.~\ref{Sec:Full_Plasma_Model}, as calculated using the algorithms developed by~\cite{Psaltis2012} (C1) and by~\cite{Younsi2016} (C2).
The secondary (lower) plots in each of the six panels show the fractional differences between the two algorithms, which are always at the $< 10^{-6}$ level.
In all panels, the dimensionless spin of the black hole is set to $a_{\sub *}=0$.
Panels in the top row correspond to radial cross-sections on the equatorial plane ($\theta=90^{\circ}$), whereas the bottom row panels correspond to radial cross-sections on a plane at a polar angle of $\theta=45^{\circ}$.
Panels in the left, middle, and right columns correspond to the Kerr metric, the JP metric with $\alpha_{\sub 13}=2$, and the JP metric with $\epsilon_{\sub 3}=2$, respectively.
All other deviation parameters are zero.
In all panels, vertical dashed lines delineate the spherical radii ($r$) or cylindrical radii ($\varpi$) of the event horizon ($r_{\sub H}$), of the UPO ($r_{\sub UPO}$), and of the ISCO ($r_{\sub ISCO}$).
The leftmost shaded grey region in each panel denotes the region interior to the event horizon of the black hole.
The relevant plasma velocity parameters are $\eta = 0.5$ and $n_{r} = 1.5$.}
\label{fig:CC4vel_a0}
\end{figure*}

\begin{figure*}[htb!]
\begin{center}
\includegraphics[width=0.34\textwidth]{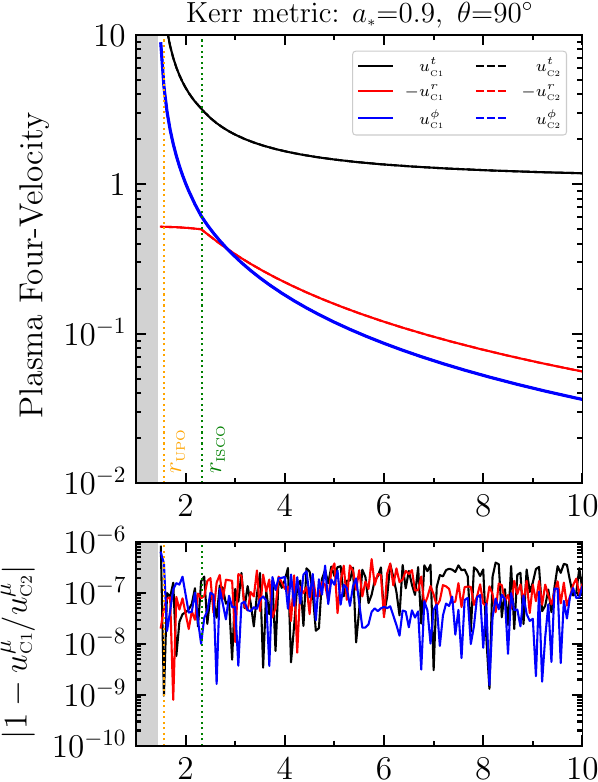}\hfill
\includegraphics[width=0.31\textwidth]{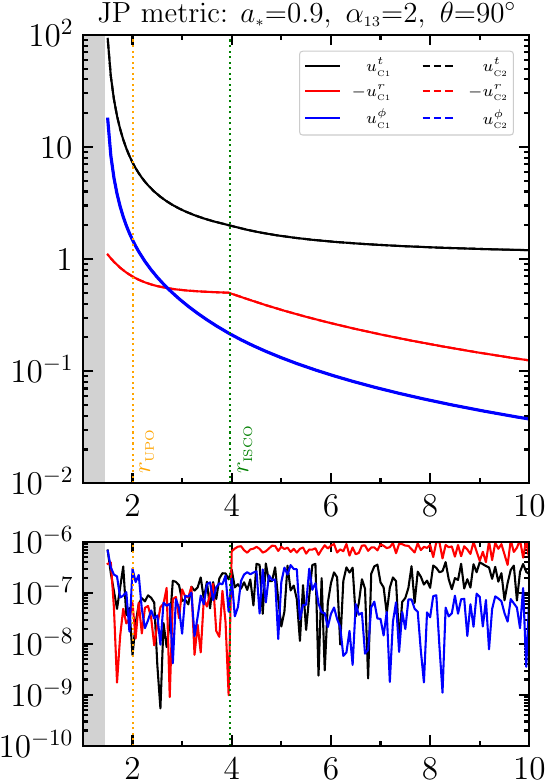}\hfill
\includegraphics[width=0.31\textwidth]{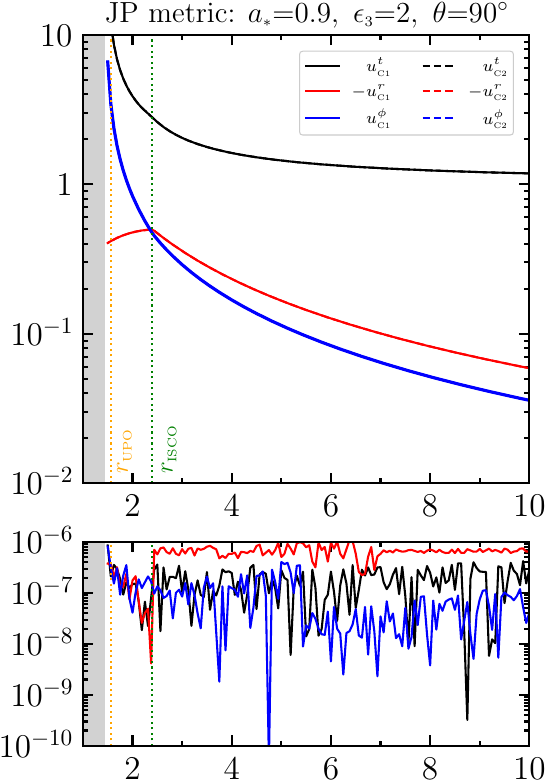}\hfill
\end{center}
\vspace*{-3mm}
\begin{center}
\includegraphics[width=0.34\textwidth]{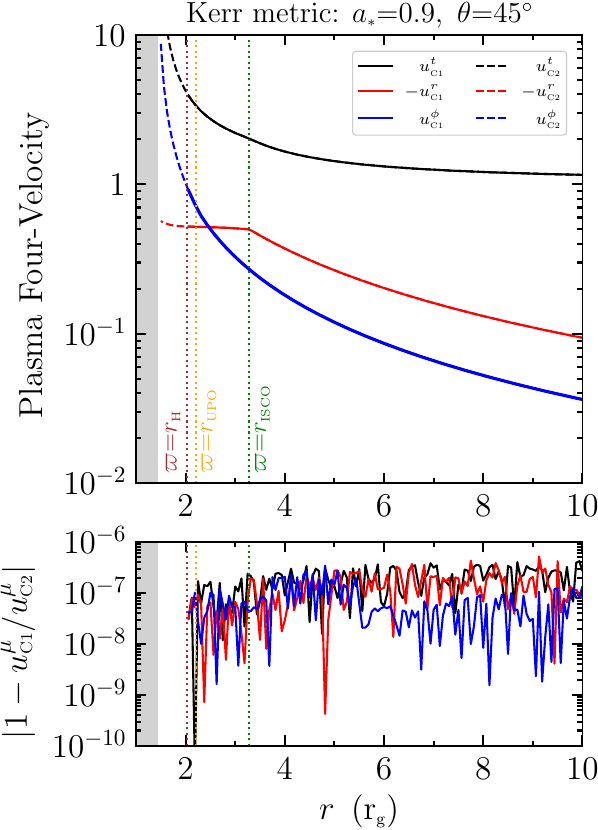}\hfill
\includegraphics[width=0.31\textwidth]{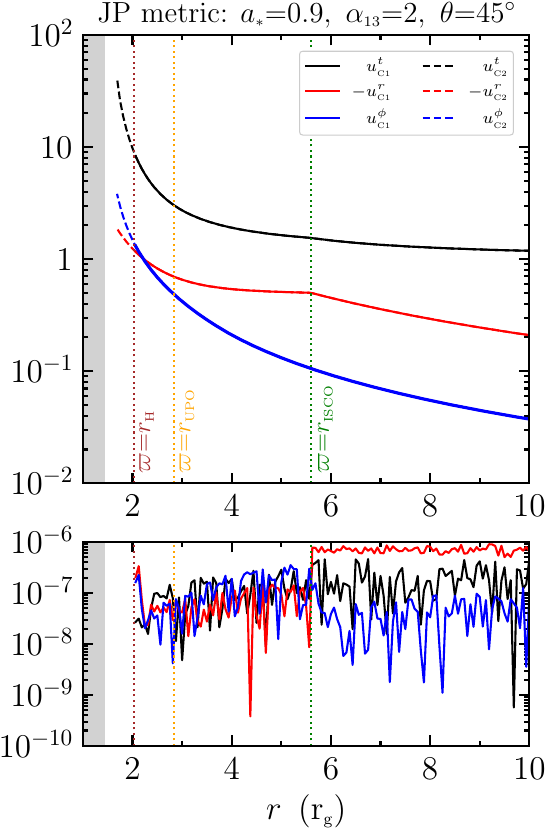}\hfill
\includegraphics[width=0.31\textwidth]{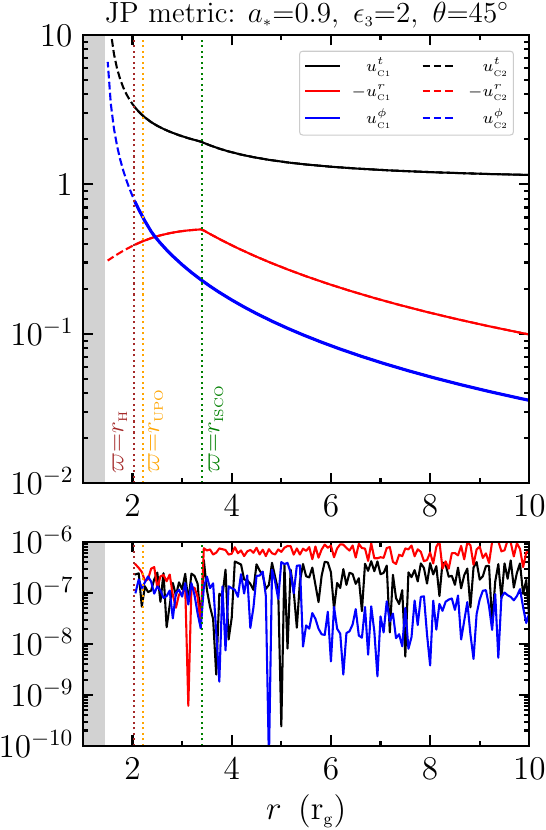}\hfill
\end{center}\vspace*{-3mm}
\caption{\footnotesize Same as Figure~\ref{fig:CC4vel_a0} but for black holes with a dimensionless spin parameter of $a_{\sub *}=0.9$.}
\label{fig:CC4vel_a0.9}
\end{figure*}

Figures~\ref{fig:CC4vel_a0} and \ref{fig:CC4vel_a0.9} compare the profiles of the three non-zero components of the plasma four-velocities calculated as described in Sec.~\ref{Sec:full_plasma_4_vel}.
These velocity components depend on different combinations of metric elements and their derivatives throughout the spacetime, as well as on the location of the ISCO, which is calculated, in general, numerically.
These profiles, therefore, provide an intricate verification of the different spacetime metric implementations.
These figures present comparisons for the Kerr and JP metrics, for different black-hole spins, and along different radial cross-sections of the three-dimensional (axisymmetric) domain.
In all cases, the fractional difference between the two algorithms is $< 10^{-6}$, which is the target accuracy imposed in the numerical calculation of the characteristic radii.

\begin{figure*}[htb!]
\begin{center}
\includegraphics[width=1.0\textwidth]{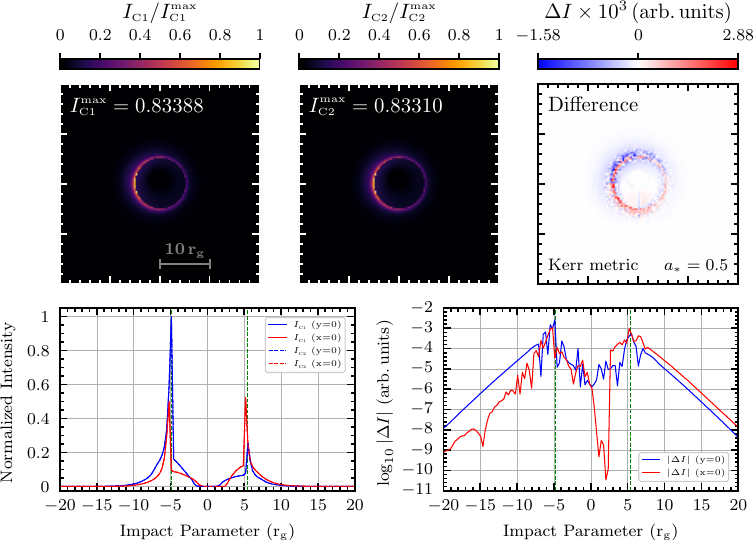}
\end{center}\vspace*{-3mm}
\caption{\footnotesize Comparison of $1.3$~mm images of a Kerr black hole with dimensionless spin parameter $a_{\sub *}=0.5$ and observer inclination angle $i=15^{\circ}$, as calculated using the algorithms developed by~\cite{Psaltis2012} (C1) and by~\cite{Younsi2016} (C2).
The upper panels show the two images as well as their pixel-by-pixel difference, and the field of view is $[-20~\rg,\, 20~\rg]$ in both directions.
The bottom-left panel shows a horizontal and a vertical cross-section of the images from both codes, with the bottom-right panel showing their difference.
The vertical green dashed lines in the bottom panels correspond to the locations of the left and right horizontal critical impact parameters.
The two algorithms generate images which agree to the $\lesssim 10^{-3}$ level.
}
\label{fig:CC_full_plas_GR}
\end{figure*}

\begin{figure*}[htb!]
\begin{center}
\includegraphics[width=1.0\textwidth]{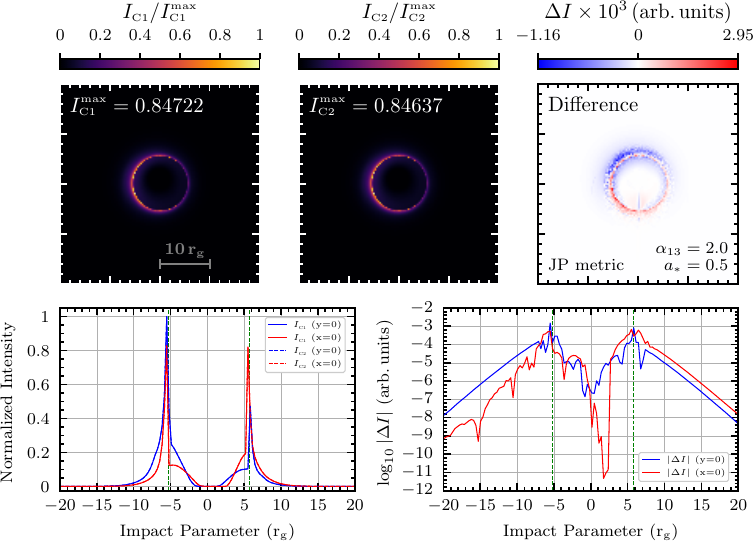}
\end{center}\vspace*{-3mm}
\caption{\footnotesize Same as Figure~\ref{fig:CC_full_plas_GR} but for a black hole described by the JP metric, with $\alpha_{\sub 13}=2.0$ the only non-zero deviation parameter.}
\label{fig:CC_full_plas_JP}
\end{figure*}

Geodesic integration is performed to an accuracy better than $10^{-12}$, with the discrepancy between both codes in evaluating the synchrotron emissivity being $\sim10^{-16}$ across all parameter values.
Consequently, it has been established that the leading source of discrepancy between codes arises from the accuracy by which the characteristic radii are evaluated, particularly the ISCO radius.
Figures~\ref{fig:CC_full_plas_GR} and \ref{fig:CC_full_plas_JP} compare the images calculated with the two algorithms for the plasma model described in \S3, for the Kerr and the JP spacetimes.
This comparison verifies the implementation of the integration of null geodesics as well as of the thermodynamic plasma quantities and the synchrotron emissivities.
The fractional difference between the images calculated with the two algorithms is larger for impact parameters that graze the photon orbits in both spacetimes.
This is expected given the large gradients in the intensity near these impact parameters.
Nevertheless, in all cases the fractional difference is $\lesssim 10^{-3}$, which is more than adequate for the purposes of the calculations reported here.

\clearpage

\bibliography{shadows_metric.bib,shadows_early.bib}

\end{document}